\documentclass[12pt,english]{scrartcl}

 \usepackage{bezier}
 \usepackage{latexsym}
 \usepackage[english]{babel}
 \usepackage{graphicx}
 \usepackage{amsmath}
 \usepackage{amssymb}
 \usepackage[ps2pdf,urlcolor=blue,colorlinks=true]{hyperref}

\newcommand{\be}{\begin{equation}}
\newcommand{\ee}{\end{equation}}

\begin{document}

\title{2-d Microcavities:  Theory and Experiments
\thanks{Contribution for \textit{Cavity-Enhanced Spectroscopies},
  edited by Roger D. van Zee and John P. Looney (a volume of
  Experimental Methods in the Physical Sciences), Academic Press, San Diego, 2002} }

\author{
  Jens U.~N{\"o}ckel\\ 
Department of Physics\\
University of Oregon, Eugene, OR 97403-1274\\
\url{darkwing.uoregon.edu/~noeckel}
  \and
Richard K.~Chang\\
Department of Applied Physics\\
Yale University, 
New Haven, CT 06520\\
\url{http://www.eng.yale.edu/rkclab/home.htm}
}


\maketitle
\begin{abstract}
An overview is provided over the physics of dielectric microcavities
with non-paraxial mode structure; examples are microdroplets and
edge-emitting semiconductor microlasers. Particular attention is given
to cavities in which two spatial degrees of freedom are coupled via the
boundary geometry. This generally necessitates numerical computations
to obtain the electromagnetic cavity fields, and hence intuitive
understanding becomes difficult. However, as in paraxial optics, the 
ray picture shows explanatory and predictive strength that can guide
the design of microcavities. To understand the ray-wave connection 
in such asymmetric resonant cavities, methods from chaotic
dynamics are required. 
\end{abstract}
\newpage
\tableofcontents
\newpage

\section{Introduction}
Maxwell's equations of electrodynamics exemplify 
how the beauty of a theory is captured in the formal simplicity 
of its fundamental equations. Precisely for this reason, they also 
illustrate that physical insight cannot be gleaned from the 
defining equations of a theory {\em per se} unless we understand how 
these equations are solved in practice. In optics, one powerful 
approach to this challenge is the {\em short-wavelength approximation} 
leading to the ray picture. 
Rays are the solutions to Fermat's variational principle, which in 
particular implies the laws of reflection and refraction at dielectric 
interfaces. As soon as one makes the transition from wave physics to 
this classical domain, concepts such as ``trajectory'', ``phase space'' 
and ``diffusion'' become meaningful. In this chapter, special attention 
will be devoted to the classical phenomenon of {\em chaos} in the ray 
dynamics of small optical cavities, i.e., a 
sensitive dependence of the ray traces on initial conditions. The 
main implication for the corresponding wave equation is that 
it cannot be reduced to a set of mutually independent ordinary 
differential equations, e.g., by separation of variables. Even when 
only a few degrees of freedom are present in the system, their coupling 
then makes it impossible to label the wave solutions by a complete set 
of ``quantum numbers'' (e.g., longitudinal and transverse mode 
numbers). Such a problem is called non-integrable; this 
classification is due to Poincar{\'e} and applies both to rays and 
waves \cite{reichl}. Integrable systems are characterized by a 
complete set of ``good quantum numbers'', as will be illustrated in 
section \ref{sec:integrableshapes}

Microcavities can be broadly categorized into active (light 
emitting) and passive systems. In {\em active} devices, such as lasers, 
radiation is generated within the cavity medium, without any incident 
radiation at the same frequency; the cavity provides feedback and (in 
combination with a gain medium) sets the emission wavelength 
\cite{siegman}. Some of the important properties that characterize a 
light emitting cavity are emission spectra, emission directivity, 
output energy, noise properties and lasing thresholds. 
Examples for {\em passive} systems are filters, multiplexers and other 
wavelength-selective optical components; their functionality relies on 
coupling to waveguides or freely propagating waves in their vicinity.
To characterize and operate passive cavities, one is interested in 
transmission and reflection coefficients, i.e., in the resonator as 
a scatterer. 

The two basic types of experiment, 
emission and scattering, for a given cavity geometry are 
intimately connected because they both probe its {\em mode 
structure}. The concept of {\em modes} in microcavities will be made 
more precise in section \ref{sec:resoqbstates}; as a working 
definition, let us denote as modes all electromagnetic excitations 
of the cavity which show up as peak structure in scattering or 
emission spectra and are caused by constructive interference in the 
resonator. 

Nonintegrable cavities introduce added freedom into the 
design of novel optical components, especially when we apply results
from the field of {\em quantum chaos}, in which 
modern quasiclassical methods are of central importance 
\cite{gutzwillerreview,
tomsovicheller,gutzwillerbook,child,brack,smilansky,stoeckmann}. 
The term ``quantum'' in ``quantum chaos'' relates to 
the fact that cavity modes are discrete as a consequence of the 
constructive interference requirement mentioned above; 
the term ``chaos'' refers to a degree of complexity in the ray 
picture which renders inapplicable all simple  
quantization schemes such as the paraxial method.

The optics problem we are addressing is not one of ``quantum 
optics'', but of the quantized (i.e., discrete) states of classical 
electrodynamics in spatially confined media. 
{\em By quasiclassics, we therefore mean the 
short-wavelength treatment of the classical electromagnetic field}. 
Confusion should be avoided between quasiclassics as defined here, 
and the semiclassical treatment of light in the matter-field 
interaction, which couples quantum particles to the electromagnetic 
field via the classical (vector) potentials: how we {\em obtain} 
the modes of a cavity (e.g., quasiclassically), should be 
distinguished from how we {\em use} them (e.g., as a basis in
quantum optical calculations). Our main emphasis in 
this chapter will be on the ``how'' part of the problem, but the 
wealth of physics contained in these modes themselves 
points to novel applications as well. Having made this clarification, 
we henceforth use the term ``semiclassics'' synonymously with 
``quasiclassics'' as defined above. 

\section{Dielectric microcavities as high-quality resonators}
\label{sec:dielectrics}
Many of the microcavities which are the subject of this chapter employ 
``mirrors'' of a simple but efficient type: totally 
reflecting abrupt dielectric interfaces between a dielectric body and 
a surrounding lower-index medium. Resonators occurring in nature, such 
as droplets or microcrystals, use this mechanism to trap 
light. This allows us to draw parallels between nature and a variety 
of technologically relevant resonator designs that are based on the 
same confinement principle. A recurring theme in numerous systems is 
the combination of {\em total internal reflection} (TIR) 
with a special type of 
internal trajectory that skips along the boundary close to grazing 
incidence: the ``whispering-gallery'' (WG) phenomenon, named after an
acoustic analogue in which sound propagates close to the curved walls 
of a circular hall without being audible in its center 
\cite{rayleigh,anderson}. An example for 
the use of the WG effect in {\em cavity ring-down spectroscopy} is reported
in Ref.\ \cite{ringdown}. A WG cavity can provide the low loss needed to 
reduce noise and improve resolution in the detection of trace species;
the trace chemicals are located 
outside the cavity and couple to the internal field by {\em 
frustrated total internal reflection}, or evanescent fields. 
In this section, we review Fresnel's 
formulas before introducing the WG modes. This provides the basis for 
sections \ref{sec:integrableshapes} - \ref{sec:experimentdynec}

In the ray picture, the Fresnel reflectivity, $R_{Fresnel}$, 
of a dielectric interface 
depends on the angle of incidence $\chi$ (which we 
measure with respect to the surface normal) and relative 
refractive index $n$. In 
particular, $R_{Fresnel}$ drops significantly below 
the critical angle for total internal reflection, 
\begin{equation}\label{eq:criticalangle}
\chi_{\rm TIR}\equiv\arcsin\frac{1}{n}.
\end{equation}
In the limit of normal incidence on a plane dielectric interface, 
the reflectivity becomes independent of polarization,
\begin{equation}\label{eq:fresnelreflchi0}
R_{Fresnel}=\left(\frac{n-1}{n+1}\right)^2.
\end{equation}
If the electric field is polarized perpendicular to the plane of 
incidence (we shall denote this as transverse magnetic, TM,
polarization), then Eq.\ (\ref{eq:fresnelreflchi0}) constitutes 
the lower bound for the reflectivity. Hence, the limit of a 
closed cavity is  approached with increasing refractive index of 
the cavity. If, on the other hand, the electric field lies in the 
plane of incidence (TE polarization), then the reflectivity 
drops to zero at the {\em Brewster angle},
\begin{equation}\label{eq:brewsterangle}
\chi_B=\arcsin\frac{1}{\sqrt{1+n^2}},
\end{equation}
and thus the closed-cavity limit is never fully reached. 
Because $\chi_{B}<\chi_{\rm TIR}$, the polarization does not affect
the trapping of  
light at incident angles $\chi>\chi_{\rm TIR}$ (to leading order in 
frequency). In addition, because 
the wavelength $\lambda$ does not enter the Fresnel formulas, internal 
reflection can be classified as a ``classical'' phenomenon which can 
be understood based on Fermat's principle. 

This should be contrasted 
with the intrinsic frequency dependence of Bragg reflection -- the
other widespread mechanism for confining light in
dielectrics. From an engineering point of view, Bragg reflectors are
challenging to realize for lateral confinement. In particular for low-index
materials, high-reflectivity windows (stop bands) are direction- {\em
and} frequency dependent with narrow bandwidth. TIR, on the other hand, is 
broad-band and technologically simple. 

At higher orders of $\lambda$, wavelength-dependent corrections to
Fresnel's formulas do arise because TIR is truly 
total only for a plane wave incident on an infinite and flat 
interface. The latter does not 
hold for boundaries with finite curvature or even sharp corners, and
similarly in cases where the incident beam has curved wavefronts, as in 
a Gaussian beam \cite{johnson, mcbook,snyderlove,laeri}. The physical
reason for radiation leakage in these situations is that light {\em
penetrates} the dielectric interface to some distance which depends
exponentially on $\lambda$, as in quantum-mechanical tunneling. This
allows coupling to the radiation field outside the cavity 
\cite{johnson,mcbook}. 

The analogy to quantum mechanics rests on the similarity between the
Schr{\"o}dinger and the Helmholtz equation for the field $\psi$, 
\begin{equation}\label{eq:helmholtz}
\nabla^2\psi+n^2({\bf r})k^2\psi=0
\end{equation}
to which Maxwell's equations often reduce. Here, $n({\bf r})$ is the
index profile and $k$ the free-space wavenumber. The WGMs of a circular 
resonator are straightforward to calculate exactly, because the 
Helmholtz equation separates in cylinder coordinates $r,\,\phi,\,z$. 
The circular cylinder and sphere are the two main 
representatives of this small class \cite{brack,kerker} of integrable 
dielectric scattering problems. 
The analytic solution, due to Lorentz and Mie, of Maxwell's 
vector wave equations for a dielectric sphere, has a long history
\cite{kerker,vdhulst,nussenzveig,barberhillbook,johnson,mcbook}). 
It finds application in a wide range of different optical processes, 
ranging from 
elastic scattering by droplets to nonlinear optics \cite{rkc} and to 
cavity quantum electrodynamics \cite{haroche,kimble,kurizki}. 
Similarly, dielectric cylinders are used, e.g., as models for 
atmospheric ice particles \cite{schmitt} or edge-emitting microdisk and 
micropillar lasers \cite{slusher,seongsikchang,mls}. 
Because of rotational symmetry, the ray 
motion in a sphere is always confined to a fixed plane. By contrast, 
rays in a cylinder can spiral along the axis
\cite{schweigercylinder,poon},  
but the propagation becomes planar 
if the incident wave is aligned to be perpendicular to the 
cylinder axis. Similar symmetry arguments make it possible to find 
analytic solutions in concentrically layered dielectrics or ring 
resonators \cite{artemyev,mazumder,mkchin}.

Each time a degree of freedom can be separated due to symmetry, 
the effective dimensionality of the remaining wave equation is 
reduced by one -- in the above examples one finally arrives at an 
ordinary (one dimensional) equation for the radial coordinate. 
Even without symmetry properties such as in the sphere or 
cylinder, one often finds approximate treatments 
by which such a reduction from three to fewer 
dimensions can be justified: in fact, in 
{\em integrated optics}, most functions are performed by planar optical 
devices, for which the mode profile in the vertical direction can 
be approximately separated 
from the wave equation in the horizontal plane, 
leaving a two-dimensional problem. For 
a discussion of methods exploiting this assumption (e.g. the 
effective-index method), the reader is referred to the literature
\cite{eim}.

One of the advantages of reducing the cavity problem to two degrees 
of freedom is that polarizations can approximately be decoupled into 
TE and TM. The resulting wave 
equations are then scalar, with the polarization information residing 
in the continuity conditions imposed on the fields at dielectric 
interfaces. Therefore, the fields are formally obtained as solutions 
of Eq.\ (\ref{eq:helmholtz}) 
in two dimensions. This scalar problem is the starting point for 
our analysis. In the absence of absorption or amplification, 
the index $n({\bf r})$ which defines our 
microcavity is real-valued, and approaches a constant (taken 
here as $n=1$ for air) outside some finite three-dimensional domain 
corresponding to the cavity. 

\section{Whispering-gallery modes}\label{sec:mie}
\begin{figure}[t]
            \centering
        \includegraphics[width=11 cm]{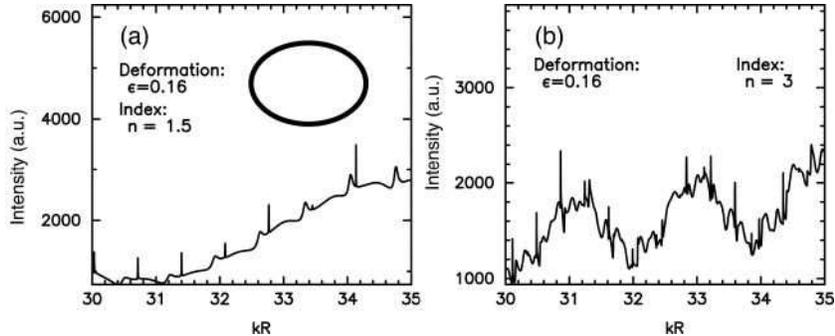}
        \caption{
Light scattering spectra for a resonator in the shape of an 
ellipse with eccentricity $e=0.8$, defined in terms of the major and 
minor axes $a$, $b$ as $e=\sqrt{1-(b/a)^2}$. 
The length scale $R\equiv\sqrt{ab}$ is used 
in our calculations to convert the wavenumber into the dimensionless 
``size parameter'' $kR$. The index of refraction 
is $n=1.5$ in (a) and $n=3$ in (b), showing how the 
openness of a dielectric cavity increases when $n$ is reduced.
The incoming plane wave in (a) and (b) travels parallel 
to the major axis, and the scattered intensity is detected at $90^{\circ}$ 
from incidence.}
        \label{fig:ellipsespec}
\end{figure}
Figure \ref{fig:ellipsespec} illustrates the formation of
whispering-gallery modes (WGMs) for a dielectric ellipse illuminated by a TM 
polarized plane wave. The wavenumber $k$ is made dimensionless by 
multiplying with the mean radius $R$ of the two-dimensional shape. 
The sharp features in the spectra of Fig.\ \ref{fig:ellipsespec} are 
caused by modes which rely on Fresnel reflection. This can be deduced
from the fact that the spectrum becomes more crowded as the refractive
index is doubled from $n=1.5$ (a) to $n=3$ (b); 
the intuitive reasoning is that 
a larger set of ray paths acquires high reflectivity as $n$ increases.
The more regularly spaced peaks of Fig.\ \ref{fig:ellipsespec} (a) can
in fact be identified as WGMs: 
Light circumnavigates the perimeter of the cavity in such a way 
as to insure TIR during a complete round-trip, 
i.e., $\chi>\chi_{\rm TIR}$ at all encounters with the boundary. 
Consequently, 
long-lived cavity modes can form by constructive interference. 
In the stability diagram of linear 
paraxial resonator theory \cite{siegman}, a circular dielectric WG 
cavity could be classified as a confocal resonator with closed but 
leaky mirrors. Gaussian-beam paraxial optics fails in circular cavities 
because the beam parameters become undefined in the confocal limit. 

\begin{figure}[t]
            \centering
        \includegraphics[width=9 cm]{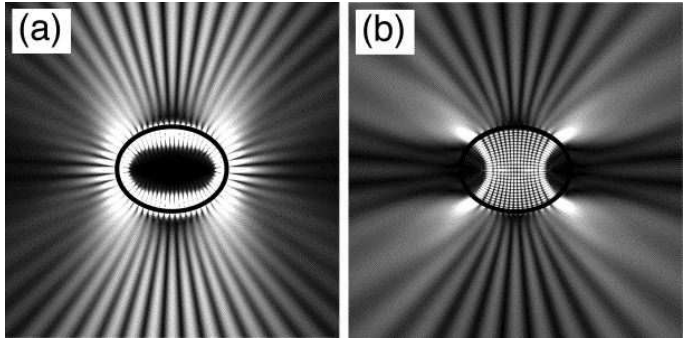}
        \caption{Numerical solution of Maxwell's equations for the 
        internal and external 
        intensity of TM modes in an elliptic cylinder with a 
        refractive index of $n=1.5$ (a) and $n=3$ (b), respectively 
        (assuming $n=1$ on the outside). 
        The eccentricity is $e=0.709$ in both plots.  
        The dimensionless wavenumber 
        in (a) is $kR=21.27636$ at a resonance width of $\kappa R=0.0085$; 
        the state in (b) is found at $kR=20.69345$ and has 
        a width of $\kappa R=0.05$. The grayscale shows high intensity 
        as white. }
        \label{fig:ellipsewaves}
    \end{figure}
The Mie treatment of circular and spherical micro-objects may serve 
as starting points for {\em perturbative} treatments of resonators 
whose shape deviates only slightly from rotational symmetry 
\cite{leung}, or when objects in the vicinity of the cavity exert a
weak influence \cite{hlwang}. 
Perturbation methods are among the most powerful tools of wave 
physics, but one must be aware that there are phenomena outside their 
reach (most apparent when energy denominators diverge). As a simple 
example, we return to the 
elliptical resonator \cite{wiersig}. If we assume impenetrable
boundaries (of Dirichlet type), it is a classic 
textbook problem \cite{landau} to obtain the WGMs of the 
ellipse by applying perturbation theory to a circular cavity. 
The weakness of this approach is revealed when we recall that 
the modes can be found by an exact separation of variables and fall 
into two classes: WGMs and beam-like ``bouncing-ball'' states; 
in Fig.\ \ref{fig:ellipsewaves} we 
illustrate these two classes with slightly more complicated 
dielectric boundary conditions. Since bouncing-ball modes do not 
{\em exist} in the circle, perturbative expansions for this type of 
modes can be expected to become problematic, especially when 
attempting to describe how some modes of the circle lose their WG 
character under a continuous shape deformation.

The wave solutions shown in Fig.\ \ref{fig:ellipsewaves} show field 
intensity extending to the exterior of the cavity because the 
dielectric interface is ``leaky''. In general, such a calculation 
must be performed numerically; we shall discuss the definition and 
emission properties of these {\em leaky modes} further below, in
section \ref{sec:qbstates} For now, we 
are only concerned with the internal intensity patterns. 
The reader will recognize a strong similarity between the 
``bouncing-ball'' mode and a higher-order transverse Gauss-Hermite 
beam; this arises because the ray motion corresponding to this mode is a 
stable oscillation between the flat sides of the cavity. The analytic 
transverse form of the bouncing-ball beam in the ellipse is however not a 
Gaussian, but a Mathieu function. The caption of 
Fig.\ \ref{fig:ellipsewaves} gives the {\em linewidth} 
$\kappa$ of the two types of modes, indicating that the WG resonance is 
almost an order of magnitude narrower than the bouncing-ball mode, 
{\em despite} the twofold higher refractive index used in 
Fig.\ \ref{fig:ellipsewaves} (b). 

The ellipse has been chosen as an example because it is integrable in 
the limit of no leakage. Other oval deformations of the circle do not 
have this simplifying property. However, families of WG ray 
trajectories have been proven \cite{lazutkinbook} to exist in 
any sufficiently smooth and oval enclosure, provided that its 
curvature is nowhere zero. This makes WGMs 
a very robust phenomenon of general convex oval cavities. 
Before we discuss in more detail the relation between the ray picture 
and the internal structure of the resonator modes, we now turn to 
some general considerations on what allows us to define the modes of 
an open cavity.

\section{Scattering resonances and quasibound 
states}\label{sec:qbstates}
WGMs are not infinitely long-lived even in an 
ideal dielectric cavity, as we saw in the finite linewidths of 
Fig.\ \ref{fig:ellipsespec}. In fact, for a finite, 
three-dimensionally confined dielectric body, all solutions to 
Eq.\ (\ref{eq:helmholtz}) are extended to infinity,
forming a continuous spectrum. A basis of eigenfunctions is given 
by the {\em scattering 
states}, consisting of an incoming wave $\psi_{{in}}$ that is 
elastically scattered by the dielectric microstructure of index 
$n({\bf r})$ into 
an outgoing wave $\psi_{out}$. In the asymptotic region where $n=1$, 
the relation between incoming and outgoing waves is mediated by the 
{\em scattering operator} $S$, known from quantum scattering 
theory \cite{weisskopf}, to which the electromagnetic resonator
problem is conceptually analogous. The S-matrix formalism has long 
been in use in microwave technology as well as optics \cite{yarivscatter}: 
A matrix representation is obtained by 
defining basis states $|\nu\rangle$ ($|\mu\rangle$) in which the 
asymptotic incoming (resp., outgoing) fields can be expanded; one has 
\begin{eqnarray}
\psi_{scat}&=&\psi_{in}+\psi_{out},\nonumber\\
\langle\mu|\psi_{out}\rangle &=&\sum\limits_{\nu=1}^{M}S_{\mu\nu}\,
\langle\nu\vert\psi_{in}\rangle .\label{eq:scatmatrix}
\end{eqnarray}
Here, $S$ is a unitary matrix if the incoming and outgoing 
waves are normalized to carry unit flux, and if $n({\bf r})$ is real. 
The dimension $M$ is the number of open scattering channels; in the 
Fabry-Perot cavity there are two channels: both, the incident field 
$|n\rangle$ and the outgoing states $|m\rangle$ can be a plane wave 
either traveling on the {\em left} or on the {\em right} of the cavity. 
In 2D and 3D, it is often useful to use angular momentum eigenstates 
as the basis defining the channels. 

Only the properties of the index profile, not of the particular 
incoming wave, enter $S$. For a review of scattering theory see, 
e.g., \cite{taylor}. As was seen in Fig.\ \ref{fig:ellipsespec}, 
the actual {\em cavity modes} in this 
continuum of scattering states are revealed if we measure the 
scattering of light as a function of wavenumber $k$. The amplitude of 
the scattered field shows resonant structure at discrete values of $k$ 
which do not depend on the detailed spatial form of the exciting field 
$\psi_{in}$. These resonances are caused by poles of $S$ in the 
complex $k$ plane, and are a characteristic of the microcavity itself. 
In Eq.\ (\ref{eq:scatmatrix}), we note that a pole of $S$ admits 
nonvanishing $\psi_{out}$ in the absence of any incoming waves, 
$\psi_{in}=0$. In these solutions at complex $k$, also known as
the quasibound states, we have finally found a proper definition of 
what we simply called ``cavity modes'' earlier. 

A well-known example is the linear two-mirror (Fabry-Perot) cavity 
\cite{siegman}. 
Its resonances are easily obtained within physical 
optics by writing the transmission 
of an incident ray as a geometric series over multiple {\em round trips}. 
In each round trip, the amplitude of a ray in the cavity accrues a factor
$
r_1r_2\,\exp(i\,n\,k\,\ell+i\,\phi_1+i\,\phi_2),
$
where $\ell$ is the round-trip path length. We have split the 
amplitude reflectivities at the individual 
reflections $\nu=1,\,2$ into modulus $r_{\nu}$ and phase $\phi_{\nu}$.  
Attenuation comes from $r_{\nu}<1$ ($\nu=1,\,2$ in the two-mirror case). 
The refractive index $n$ in the cavity is real, as stated below 
Eq.\ (\ref{eq:helmholtz}), and the wave number $k$ is measured in
free space. The 
transmitted amplitude $T_{total}$ is obtained by summing this over all 
repetitions, 
\be\label{eq:raysum}
T_{total}=I\,\sum\limits_{\sigma=1}^{\infty}\left(r_1\,r_2\,
e^{i\,n\,k\,\ell+i\,\phi_1+i\,\phi_2}\right)^{\sigma},
\ee
where the prefactor $I$ is determined by the incident wave and the 
mirror transmittivities. This geometric series can be summed, leading
to a denominator $1-{\tilde g}_{\rm rt}({\tilde\omega})$, with the 
round-trip ``gain''
\be\label{eq:roundtrip}
{\tilde g}_{\rm rt}({\tilde\omega})=e^{in{\tilde \omega}\ell/c}
\,\prod_{\nu}r_{\nu}\,e^{i\phi_{\nu}}.
\ee
In our example, $\nu$ runs from $1$ to $2$. 
Whenever ${\tilde g}_{\rm rt}$ comes close to $1$, the resonator 
exhibits a transmission peak. The equality ${\tilde g}_{\rm 
rt}({\tilde \omega})=1$ can be satisfied only if we admit complex frequencies,
${\tilde \omega}\equiv\omega+i\gamma$, and choose for the 
imaginary part 
\be\label{eq:gammareflect}
\gamma=-\frac{c}{\ell}\,\sum\limits_{\nu}\ln r_{\nu}.
\ee
On the real frequency axis, $\gamma$ determines the resonance 
linewidth. Were we to look at the complex-frequency 
solution directly and reinstate the time dependent factor 
$\exp(i\,{\tilde \omega} t)$ that accompanied the original wave 
equation, the field at every point in space would decay with a 
factor $\exp(-\gamma t)$. An interpretation of this can be given in 
the ray picture: for a ray launched {\em inside} the cavity, the 
field is attenuated by a 
factor $r_{1}\,r_{2}$ in each round trip; after time $t$, the 
number of round trips is $ct/\ell$, leading to the exponential law 
\be\label{eq:raydecay}
E(t)=e^{ct\ln(\Pi r_{\nu})/\ell}=e^{-\gamma t}.
\ee
Hence, each transmission peak is uniquely 
associated with a metastable, or {\em quasibound} state, and the decay 
rate is determined by the reflectivities encountered during a round 
trip. 

Equation (\ref{eq:raydecay}) straightforwardly justifies the concept 
of a metastable state within the ray picture. We introduced 
resonances as peaks in the transmitted field, arising as 
near-divergences of a geometric series over ray paths undergoing 
multiple reflections inside the cavity. This is in fact just a 
special case of a very general quasiclassical approach which has found 
widespread use in chemical physics \cite{eckhardtrev} 
and other fields \cite{blumelsmilansky} since its original 
introduction by Miller \cite{child,millerchem}. 
It relates the $S$-matrix 
element between arbitrary channels $\mu$ and $\nu$ of a multichannel 
scattering process to a sum over all possible ray paths starting in 
channel $\nu$ and ending in $\mu$:
\be\label{eq:miller}
S_{\mu\nu}=\sum\limits_{\sigma}\sqrt{p_{\mu\nu}^{(\sigma)}}\,\exp\left(
ik\,\Phi^{(\sigma)}-i\alpha^{(\sigma)}\right),
\ee
where $\sigma$ parameterizes the family of ray trajectories leading from 
incoming channel $\nu$ to outgoing channel $\mu$, just as 
in Eq.\ (\ref{eq:raysum}). $\alpha^{(\sigma)}$ is a 
phase shift acquired by the rays as they encounter caustics and 
interface reflections 
along their path. The phase shift $k\Phi^{(\sigma)}$ is the generalization 
of the 
dynamical phase $k\,\ell$ in the linear example, and 
$p_{\mu\nu}^{(\sigma)}$ is 
the transition probability, corresponding to the product of 
reflection and transmission coefficients in our two-mirror 
example. For a cavity defined by Fresnel reflection, 
$p_{\mu\nu}^{(\sigma)}$ 
can to lowest approximation be determined purely within ray 
optics. 

The virtue of Eq.\ (\ref{eq:miller}) is that it points the way from 
geometric optics to wave optics even in systems where the ray 
paths are not as easily enumerated as in the Fabry-Perot cavity. 
The Fabry-Perot cavity discussed above is an 
example where Eq.\ (\ref{eq:miller}) in fact yields exact results, 
because only plane-wave propagation is involved. Although 
corrections to this quasiclassical formula are necessary in more 
complicated cavity geometries, Eq.\ (\ref{eq:miller}) makes it 
plausible that ray considerations are a powerful tool for 
understanding quasibound states in many open systems. 
In this spirit of ray-based scattering theory, we can ask how to
extract the {\em quasibound states} as poles of the physical-optics expression 
Eq.\ (\ref{eq:miller}). This means we want to generalize the 
logical transition (illustrated for the Fabry-Perot cavity) from a 
transmission amplitude determined by Eq.\ (\ref{eq:roundtrip}) to an 
internal ray loop 
with attenuation given by Eq.\ (\ref{eq:raydecay}). Thus, the original 
scattering problem should be replaced by a Monte-Carlo simulation of a 
suitable ensemble of rays {\em inside the resonator}, and 
the internal ray dynamics suffers {\em dissipation} owing to the 
openness of the cavity. Several technical problems make it difficult to
carry out this idea in a general cavity: the first question 
is what would be the proper choice of initial conditions for a ray ensemble
in a two-dimensional cavity such as the ellipse of 
Fig.\ \ref{fig:ellipsewaves}. The answer is provided by quasiclassical 
quantization conditions that put some constraint on the ray paths 
to be used in the ray calculation. Employing a strategy along these lines, 
it is possible to predict not only the decay rate of a quasibound state 
as in Eq.\ (\ref{eq:raydecay}), but also the {\em directionality} of the
emitted radiation \cite{nature}. This will be expounded in sections 
\ref{sec:poincaresections} and \ref{sec:chaoticinterference}

\section{Cavity ring-down and light emission}
We now discuss the significance of complex 
frequencies in Eqs.\ (\ref{eq:helmholtz}) and (\ref{eq:scatmatrix}). 
This will highlight the relation between quasibound states and
the observable cavity response to an external field or pump signal, 
which is of particular interest in spectroscopy. 
In Eq.\ (\ref{eq:helmholtz}), the wavenumber only appears in 
the form of a product $n\,k$, so that an imaginary part in ${\tilde 
k}\equiv k+i\kappa\equiv {\tilde \omega}/c$ can immediately be 
re-interpreted as part of a complex refractive index ${\tilde n}$
at {\em real} $k$, setting 
\be\label{eq:complexn}
n\,{\tilde k}\equiv n\,(k+i\kappa)=k\,(n+i\,n\kappa/k)\equiv {\tilde 
n}\,k.
\ee
The resulting Helmholtz equation has a real wavenumber, and thus 
describes 
the steady-state wave solutions in an amplifying medium, 
because $\kappa >0$: a plane wave 
would have the form $\exp(i\omega t-i{\tilde n}kx)$, which grows in the 
propagation direction. From Eq.\ (\ref{eq:complexn}) it follows that 
quasibound states appear naturally as approximate 
solutions for the lasing modes of a microlaser with a homogeneous gain 
medium \cite{amsterdam}. For a physical understanding of lasers 
\cite{lambschleich}, the {\em openness} of the system as contained in
the quasibound state description is essential. 

When a cavity is excited with a {\em pulse}, on the other hand, we 
are not looking for steady-state solutions but for transients. As we 
noted below Eq.\ (\ref{eq:raydecay}), quasibound states describe such 
a decay process. These non-stationary states are exploited in many 
fields, e.g. nuclear physics (where they are called ``Gamov states''), 
and their properties are well-known \cite{weinstein,moiseyev}. 
One peculiar property that may cause confusion is that they formally 
{\em diverge} in the far field, as can be seen by noting that the 
outgoing wave in Eq.\ (\ref{eq:scatmatrix}) obeys the radiation boundary 
condition, which in the continuation to complex frequency reads
\be
\psi_{{out}}\propto e^{i({\tilde \omega} t - {\tilde 
k}r)}\qquad(r\to\infty).
\ee
This holds for any dimensionality if $r$ is understood to be the 
radial coordinate, and algebraic prefactors (which depend on the 
dimension of the solid angle element) are not considered. At fixed 
$t$, the imaginary part $\kappa$ of ${\tilde k}$ causes exponential 
growth with $r$. However, grouping together the exponential 
dependences on position {\em and} time, the 
amplitude of the quasibound state is controlled by the factor
\be\label{eq:crdfactor}
e^{-\gamma t +\gamma r/c}.
\ee
Clearly, this remains constant along the spacetime trajectory 
$r=ct$. The complex-valuedness of ${\tilde\omega}$ 
${\tilde k}$ therefore has nothing unphysical, 
provided {\em causality} is taken into account: within the ring-down 
time $1/\omega$ of the cavity, the maximum distance to which we can 
extend measurements of the radiated field is of order $R\sim 
c/\gamma=1/\kappa$. In this range, $r<R$, the decaying 
pre-exponential factors dominate over the exponential $r$ dependence. 

Equation (\ref{eq:complexn}) represents a level of approximation which 
does not take microscopic properties of the light-emitting 
medium into account. However, it 
forms a starting point which emphasizes the effect of the resonator 
boundaries on the mode structure from the outset; an effect that 
becomes essential as the cavity size decreases. If we allow the 
polarization ${\bf P}$, which in Maxwell's wave equation 
effects the coupling between matter and field, to become a function of 
position, and possibly acquire a nonlinear dependence on the electric 
field, then it turns out that the quasibound states of the homogeneous 
medium, discussed above, are nevertheless a convenient basis in which 
to describe the emission and mode coupling caused by ${\bf P}$. To 
make this plausible, recall that any given electromagnetic cavity 
field can be expanded in an integral over the scattering states 
$\psi_{scat}$ of Eq.\ (\ref{eq:scatmatrix}) as a 
function of $k$; these are the ``modes of the universe'' in the 
presence of the cavity. To evaluate such an integral, one can extend the 
integration contour into the complex plane and apply the residue 
theorem. The integral is then converted to a sum over those $k$ at 
which the integrand has a pole. But the poles of the scattering states 
are just the poles of the $S$ matrix, i.e. the quasibound states. For 
details of these arguments, the reader should see 
Ref.\  \cite{weinstein}. The radiation 
from a source distribution $d({\bf r})$ is obtained directly by summing up those 
quasibound states which lie in the spectral interval of interest, 
weighted according to their overlap with the distribution 
$d({\bf r})$, and with an energy denominator that makes sharp 
resonances contribute more strongly than broad ones. 
In particular, for small cavities where the free spectral 
range is large and the linewidths are small, the emission properties 
are determined by the spatial form and temporal behavior of {\em 
individual quasibound states}.

\section{Wigner delay time and the density of states}\label{sec:wignerdos}
In spectroscopic applications, the quality $Q$ and finesse $F$ of a
cavity are important figures of merit. Conventionally, one defines
\be
Q\equiv \frac{k_r}{\kappa},\quad F\equiv\frac{{\rm FSR}}{\kappa},
\ee
where FSR is the free spectral range and $k_r$ is the wavenumber 
of the mode. 
To decide whether the linewidth of an individual resonance, as given by 
Eq.\ (\ref{eq:gammareflect}), is narrow or broad, we can choose as our 
yardstick the separation between neighboring modes. However, as seen in 
Fig.\ \ref{fig:ellipsespec} (b), this can become rather ambiguous if 
the spectrum is crowded. A free spectral range is then hard to define, and 
should be replaced by a continuous function of frequency: the spectral 
density, or {\em density of states}, $\rho(k)$. If $Q$ and $F$ were 
independent of $k$, then the total number of cavity modes with wavenumber 
less or equal to $k$ would be $N(k)=k/{\rm FSR}=Q/F$. The density of
states is the $k$-dependent generalization of the inverse FSR:
\be
N(k)=\int_0^{k}\rho(k')\,dk'.
\ee
The structure of $\rho(k)$ contains all the information about 
both $Q(k)$ and $F(k)$. 

For a dielectric scatterer with sharp boundaries, there is in 
principle a clear distinction between ``inside'' and ``outside''. An 
incoming wave propagates freely outside the dielectric, but can be 
trapped inside for some time. Therefore, an incident wave pulse 
emerges from the scatterer with a time delay, $\tau_{D}$, compared to 
the time it takes in the absence of the obstacle ($S\equiv 1$). 
The time delay is a continuous function of $k$, whereas the 
resonance decay time $1/\gamma$ labels discrete poles of the S-matrix.
To clarify the relation between {\em delay time} and {\em decay time}, 
consider the simplest case of a one-channel scattering system, in which 
$\psi_{in}\propto\exp(i\omega t+ikx)$ and 
$\psi_{out}\propto\exp[i\omega t-ikx+i\theta(k)]$. Here, $\theta$ is the 
{\em scattering phase shift}, which is related to the (unitary) 
S-matrix of Eq.\ (\ref{eq:scatmatrix}) by 
\be
S(k) = \exp[i\theta(k)].
\ee
The outgoing wavepacket then has the Fourier decomposition 
\be\label{eq:wavepacket}
\psi(x,t)=\int\!\!dk\,\xi(k)\,e^{ik(ct-x)+i\theta(k)}.
\ee
We can look at this general expression in two instructive limiting cases:
the magnitude of of the Fourier spectrum, $|\xi(k)|$, could be assumed to be
either sharply peaked or slowly-varying. Consider first the case where 
$|\xi(k)|$ has a narrow peak at some central wavenumber $k_0$; the 
opposite limit will be treated in Eq.\ (\ref{eq:broadc}). 
Then the variation of $\theta$ with $k$ 
need only be retained to linear order, yielding 
\be
\theta(k)\approx\theta(k_{0})+(k-k_{0})\,\frac{d\theta}{dk_{0}}
\quad\Rightarrow\quad
\psi(x,t)\approx {\rm const}\times
\int\!\!dk\,\xi(k)\,e^{ik(ct-x+d\theta/dk_{0})}.
\ee
If we denote by $\psi_{0}(x,t)$ the corresponding pulse for 
$\theta\equiv 0$, i.e. without scattering, then the last equation means
\be
\psi(x,t)=\psi_{0}\left(x,t+\frac{1}{c}\frac{d\theta}{dk_{0}}\right).
\ee
This shift in $t$ is the {\em Wigner-Smith delay time}, $\tau_{D}(k)$ 
\cite{smith,fyodorov}; it is a continuous function of $k$. 
For or a general scattering 
system with $M\ge 1$ channels, $\tau_{D}(k)$ is obtained by adding the 
$k$ - derivatives of all the phases $\theta_{\mu}$ entering the $M$ 
eigenvalues $\exp(i\theta_{\mu})$ of the $S$ matrix:
\be\label{eq:delaytime}
\tau_{D}(k)\equiv
\frac{1}{M\,c}\,\sum\limits_{\mu=1}^M\frac{\partial\theta_{\mu}}
{\partial k}=
\frac{i}{c}\,{\rm Tr}\left(\frac{dS^{\dagger}}{dk}S\right)
=\frac{1}{c}\,{\rm Im}\,{\rm Tr}\left(\frac{d}{dk}\ln 
S(k)\right).
\ee
The last equality is a matrix identity; it 
contains a trace over all open channels $\mu=1\ldots M$. The logarithmic 
derivative can be executed immediately if one channel is resonant: 
near a resonance frequency $\omega_{r}=ck_{r}$ one has 
\be\label{eq:resonancedenominator}
S\propto 1/(k-k_{r}-i\kappa)
\ee
with $\kappa$ describing the 
linewidth, or the decay rate in Eq.\ (\ref{eq:crdfactor}) 
via $\gamma=c\kappa$.  Then 
\be\label{eq:delayresotime}
\tau_{D}(k_{r})\approx \frac{1}{c\kappa}=\frac{1}{\gamma},
\ee
which means that the {\em resonant} time delay equals the resonance 
lifetime. 

The spectral density of the open system is closely 
related to the delay time $\tau_{D}$, as we can understand from the following 
(non-rigorous) argument (for more stringent derivations, see, e.g., 
\cite{krein}):
$\tau_{D}$ defines a {\em length scale} 
$L_{\tau}\equiv c\,\tau_{D}/n$ ($n$ is the refractive index) which is the 
characteristic distance over 
which the wave will propagate inside the scatterer. If we interpret 
this as the effective ``cavity length'', then a mode should naively 
be expected when an integer number of wavelengths fits into this 
length, i.e., 
\be
nk\,L_{\tau}=2\,\pi\,\nu+const,
\ee
where $\nu$ is an integer, and the constant takes into account phase 
shifts at interface reflections or caustics. The number of modes 
that are contained in a small interval $\Delta k$ around $k$ is 
then given by the corresponding change $\Delta \nu$ in the above 
equation, which to lowest order in $k$ is 
\be
\Delta \nu=\frac{1}{2\,\pi}\,L_{\tau}\,n\Delta k.
\ee
Therefore, the spectral density is simply
\be\label{eq:kreinfriedel}
\rho(k)=\frac{\Delta \nu}{\Delta k}=\frac{n}{2\,\pi}\,L_{\tau}=
\frac{c}{2\,\pi}\tau_{D}(k)
\ee
This, combined with Eq.\ (\ref{eq:delaytime}),
is the {\em Krein-Friedel-Loyd formula} for the density of 
states in an open system, which re-appears almost invariably whenever 
linear response or time-dependent perturbation theory in the presence 
of a continuous spectrum are 
considered \cite{thirring,friedel,krein,loyd,loydsmith,
faulkner,gaspard,wirzba1}. An important example is Fermi's golden rule 
which relates electronic transition rates $W$ to the squared matrix 
element $|M|^2$ of the interaction and the density of states,
\be\label{eq:fermirule}
W(k)=\frac{2\pi}{\hbar^2}|M|^2\rho(k).
\ee
Based on this and Eq.\ (\ref{eq:kreinfriedel}), we can state that 
microcavities are able to enhance optical transition rates, because 
at certain wavenumbers the light is trapped (i.e., delayed) in the cavity 
for long times. 

The density of states is one of the fundamental quantities that make
small cavities interesting, as was recognized long ago by Purcell 
\cite{purcell,feld,meschede,Yamamoto,brorson,laerinoeckelbook} 
who observed that Eq.\ (\ref{eq:fermirule}), when applied to the 
probability for spontaneous emission, can lead to an 
enhancement of this atomic decay process compared to its rate in free 
space, by many orders of magnitude. 
What we have seen here is that peaks in the spectral density 
are associated with rapid variations in the $S$ matrix, i.e., with 
the quasibound states discussed in section \ref{sec:qbstates}. Small
cavities also enable the opposite effect, a suppression of spontaneous
emission when no cavity modes fall within the emission spectrum 
\cite{meschede,kleppner,yablonovic}. 

The limit of a well-defined frequency in the wavepacket of 
Eq.\ (\ref{eq:wavepacket}) allowed us to assume that the radiation 
interacts only with a single quasibound state, leading to the 
resonant delay time Eq.\ (\ref{eq:delayresotime}). On the other hand, 
if we take $|\xi(k)|\equiv C$ to be constant, 
the whole spectrum enters with equal 
weight. One could still make the assumption that there is only a single 
isolated resonance in the spectrum; then one immediately arrives at
the well-known relation between Lorentzian lineshape and 
exponential resonance decay: take the resonance to be at wavenumber
$k_r$ (it is in fact always accompanied by a partner state \cite{leung} at 
$-k_r$, but we can ignore it at large enough frequencies); 
replacing $exp(i\theta)=S$ in Eq.\ (\ref{eq:wavepacket}) 
by Eq.\ (\ref{eq:resonancedenominator}), what remains is the 
Fourier transform of a Lorentzian, 
\be\label{eq:broadc}
\psi(x,t)=C\,\int\!\!dk\,\frac{1}{k-k_{r}-i\kappa}e^{ik(ct-x)}
\propto e^{-\kappa(ct-x)}e^{ik_{r}(ct-x)}.
\ee
This is a damped oscillation as a function of $ct -x$ at 
frequency $\omega_r=c\,k_r$ with decay constant $\gamma_r=c\kappa_r$. 
Comparing with Eq.\ (\ref{eq:crdfactor}), we have recovered the decay 
law of an individual quasibound state in the field envelope. 

According to Eq.\ (\ref{eq:broadc}), if we could make a time-resolved 
measurement of the oscillating
electric field in the outgoing wave, the resonance lifetime would
show up in the decaying envelope of the successive peaks of the rapidly
oscillating field. It is impractical to make such a measurement 
at optical frequencies. However, recent advances have been able to 
approach this limiting case of 
Eq.\ (\ref{eq:wavepacket}) by making the wavepackets into ultrashort 
pulses, where correspondingly $|c(k)|$ becomes very broad. This will be
discussed further in the next section. The results 
of a real experiment are, however, modified from this simple Lorentzian
approximation because the scatterer in general supports more than one
distinct quasibound state. The Fourier transform in 
Eq.\ (\ref{eq:wavepacket}) consequently does not yield a purely 
exponential envelope, except at long times when the decay is
dominated by the narrowest resonance \cite{dittes}. 

\section{Lifetime versus linewidth in experiments}
\label{sec:experiments}
The two limiting cases of $\xi(k)$ in the wavepacket 
Eq.\ (\ref{eq:wavepacket})
are just extremes of the time-frequency uncertainty
relation. Experiments on microcavities have been performed both in the
spectral and time domain. 
However, there has been only one report of the temporal nature of an
{\em ultra-short} optical wavepacket (100 fs pulse, 30 µm short in
air) incident on a pendant-shaped, hanging droplet with an equatorial
radius of 520 µm \cite{jpwolf}.  Such pulses are spectrally broad but
have the intriguing property that their spatial length is much shorter
than the cavity size. Conceptually, the analogy to a well-defined
particle trajectory suggests itself, and hence one looks for 
{\em ballistic} propagation in the cavity. In the experiment, 
wavepackets were observed to circulate inside the droplet in the
region where the the WGMs reside; such paths are characteristic of
high-order rainbows in that entry and exit of the light are separated
by a large number of internal reflections \cite{nussenzveig,lockellipse}.

The time-resolved measurement showed that 
coherent excitation of a large number of WGMs allows propagation of a
short wavepacket along the sphere's equator for several round trips without
significant decoherence, except for decrease in the wavepacket
intensity because of leakage of the WGMs.   
The novel aspect of the experiment that enabled the
time-resolved measurements was the use of two-color two-photon-excited
Coumarin 510 dye molecules embedded in the liquid (ethylene glycol)
that formed the pendant droplet.  The Coumarin fluorescence (near $510$
nm) appears when one wavepacket of $\lambda_1$ and another wavepacket 
of $\lambda_2$ spatially overlap.  Thus, the Coumarin fluorescence
acts as a correlator between wavepackets. 

This time-resolved experiment was designed to answer the
following questions:  (1) does the excitation wavepacket remain intact
and ballistic after evanescent coupling with WGMs? (2) after a few
round trips, would dispersion cause the wavepacket to broaden? and 
(3) can the cavity
ring down time be observed for those wavepackets that make several
round trips?  The answers were reached that the shape of excitation
wavepacket remained intact, the wavepacket was not broadened after a
few round trips, and that cavity ring down was observed for each
round trip. 
This is an extension of the single-mode ring-down determined by 
Eq.\ (\ref{eq:broadc}). 

In particular in the context of cavity ring-down spectroscopy, the
usefulness of time-domain measurements is recognized \cite{muertz}.
There, one deals with very high Q-factors and their modification by 
the sample to be studied. Another application of temporal
observation is encountered in microdroplets, where 
the optical feedback provided by WGMs makes it possible to
reach the threshold for stimulated Raman scattering (SRS). 
The SRS spectrum consists of sharp peaks, commensurate
with the higher $Q$ WGMs located within the Raman gain profile.  The
highest $Q$ value of the WGMs can be determined either by resolving the
narrowest linewidth $\Delta\lambda$ of the SRS peak or by measuring 
the longest exponential decay of the SRS signal after the pump laser 
pulse is off.  When $Q >10^5$, the decay time $\tau = Q/\omega$ 
(where $\omega\approx 3\times 10^{15}$ Hz for $\lambda = 620$ nm)
becomes a much easier quantity to measure directly because it is
longer than $100$ ps.   Otherwise, the spectral linewidth 
($\Delta\lambda = \lambda/Q$) needs to be
resolved better than $0.006$ nm (for $\lambda = 600$ nm). 
Cavity decay lifetimes as long as $6.5$ ns have been 
observed \cite{zhang}. 

\section{How many modes does a cavity support?}
\label{sec:resoqbstates}
Having defined the density of modes for the open system
in section \ref{sec:wignerdos}, we can go one step 
further and perform the the {\em average} ${\bar \rho}(k)$ over some 
finite spectral interval $\Delta k$. The suitable choice for such an 
averaging interval should of course contain many spectral peaks. Two 
relevant examples where an average spectral density is of use are the 
Thomas-Fermi model of the atom \cite{friedrich} and Planck's radiation 
law. The latter shall serve as a motivation for some further discussion of
the average ${\bar \rho}(k)$ in the next section. 
For a presentation of the subtleties involved in this procedure, 
cf.\ Refs.\ \cite{haake,bluemel}. For our purposes, we simply remark that
using Eq.\ (\ref{eq:kreinfriedel}) 
as a starting point, one way of arriving at an averaged spectral density 
is to make the formal substitution $k\to k+iK$. This amounts to an 
artificial broadening of all resonances as a function of $k$ to make 
them overlap into a smoothed function; $K$ then plays the role of 
the averaging interval \cite{balian}. 

In the limit of the closed cavity, the spectral density becomes a 
series of Dirac delta functions as the resonance poles move onto the 
real $k$ axis and become truly bound states. The 
resonator then defines a Hermitian eigenvalue problem of the type we 
encounter in all electromagnetics textbooks, and many fundamental 
properties of realistic cavities can be understood within this 
lossless approximation. As a point in case, it is worth 
recalling the problem of {\em blackbody radiation}. From a 
historical point of view, this thermodynamic question was the nemesis 
of classical mechanics as the foundation of physics, because it led 
to the postulate of discrete atomic energy levels. From a 
practical point of view, the blackbody background can be a source of 
noise in spectroscopic measurements, and its spectrum is modified by 
the presence or absence of a cavity. 

From an {\em electrodynamic} point of view, the central nontrivial 
aspect of Planck's problem is that the average spectrum of the blackbody can 
be observed to be {\em independent} of the cavity shape. The explanation 
of this universality rests on the average spectral density of 
cavity modes, which is found to be independent of the resonator 
geometry to leading order in frequency. Although it may be intuitively 
convincing that the shape of the enclosure should become unimportant 
when its dimensions are large compared to the wavelength \cite{yariv},
the actual proof requires a large measure of ingenuity. In a series 
of works beginning in 1913 \cite{weyl}, {\em Weyl} showed that for a closed, 
three-dimensional electromagnetic 
resonator of volume $V$, the average spectral density as a function 
of wavenumber $k$ is (including polarization) 
\begin{equation}\label{eq:weyl3d}
{\bar \rho}_{\rm Weyl}^{3D}(k)\approx \frac{k^2}{\pi^2}\,V\quad{\rm 
(in \quad 3D)}, \qquad
{\bar \rho}_{\rm Weyl}^{2D}(k)\approx\frac{k}{2\,\pi}\,A\quad{\rm (in 
\quad 2D)}.
\end{equation}
The second equation applies when, as discussed above, 
the wave equation can be reduced to scalar form and 
two degrees of freedom for a cavity of area $A$. 
The average number of modes in an interval $\Delta k$ 
then is ${\bar \rho}_{\rm Weyl}(k)\,\Delta k$ for large $k$. 
Geometric features 
other than the volume enter in this quantity only as corrections with 
lower powers of the wavenumber. These terms depend on the boundary 
conditions, surface area (or circumference) and curvature, as well as 
on the topology of the cavity \cite{kac,eckhardt}. 

As mentioned in the previous section, Eq.\ (\ref{eq:fermirule}), 
microcavities can lead to 
enhanced spontaneous emission because of their highly peaked density 
of electromagnetic modes. Although Weyl's formula is strictly valid 
only for short wavelengths, it nevertheless allows us to estimate the 
limiting size for ultrasmall cavities that should be approached if we 
want to observe such density-of-states effects: 
note that Eq.\ (\ref{eq:weyl3d}) approaches a ``quantum limit'' 
as the volume $V$ approaches $(\lambda/2)^3$ at fixed wavelength 
$\lambda$: the number of modes with wavenumber below $k=2\pi/\lambda$ 
is then 
\begin{equation}\label{eq:nweyl3d}
{\bar N}_{\rm Weyl}^{3D}(k)=\int\limits_{0}^{k}
{\bar\rho}_{\rm Weyl}^{3D}(k)\,dk
\approx \frac{\pi}{3}\approx 1,
\end{equation}
which means that only a single mode remains in the cavity. 
In reality, the ultrasmall size 
approached here is not necessarily the optimal choice, because it is 
hard to maintain good quality for small cavities. One of the best ways 
to achieve both small size and long lifetimes is by making use of WGMs 
in dielectric cavities.

The universality of Weyl's spectral density is surprising if we look at 
it with the methods of the previous chapter: the analysis for stable 
resonators \cite{siegmanchapter} shows that the 
number of modes which can be obtained within the {\em paraxial 
approximation} depends strongly on details of the cavity geometry. 
Recall the nature of the paraxial approximation; it is in fact a {\em 
short-wavelength} approximation to Maxwell's wave equations, in the 
sense that the characteristic length scale (of the cavity or 
variations in the refractive-index profile) along the propagation 
direction must be large compared to the wavelength. 
In that framework, the {\em geometric-optics} picture is therefore the 
backbone on which the mode structure of a cavity is built. 
However, paraxial modes appear 
only if {\em stable}, closed ray patterns exist in 
the cavity. What makes Weyl's formula nontrivial is that it 
does not distinguish between a cavity in which such stable orbits are 
readily available (as was the main subject of the previous chapter), 
and the extreme yet realizable case of a chaotic cavity in which no stable 
paths can be found {\em at all}, see section \ref{sec:chaos}.

Stability is a property of particular rays, not of a cavity as a
whole. To understand all the modes of a generic cavity, one has to go
beyond paraxiality. It also must be realized that 
paraxial optics is 
itself a special case of a more general approximation scheme, known as 
the {\em parabolic-equation method}
\cite{laerinoeckelbook,babic}. This name refers to 
the fact that the resulting wave equation of Schr{\"o}dinger type is 
mathematically classified as a parabolic differential equation. This 
is mentioned here because at a more abstract level, 
one can build approximations of paraxial type not only 
around rectilinear rays: e.g., for WGMs, an envelope function ansatz 
in polar coordinates, $\Psi(r,\,\phi)=\chi(r;\phi)\exp(-i\beta \phi)$ 
allows $\phi$ to become the ``propagation direction''. The question of 
whether a mode can be called paraxial or not then becomes 
dependent on the coordinate system one uses (e.g., Cartesian vs.\ 
cylindrical). A more {\em fundamental} distinction by which the cavity 
as a whole can be classified is that of integrability. Next, we 
discuss some examples for this concept. 

\section{Cavities without chaos}\label{sec:integrableshapes}
One feature that can be observed in both Fig.\ \ref{fig:ellipsewaves} 
(a) and (b) is that the modes exhibit {\em caustics} inside the 
dielectric, i.e. 
well-defined curves of high intensity which in the ray picture 
correspond to envelopes at which the rays are tangent. In the WGM, 
the caustic is an ellipse {\em confocal} with the boundary; 
it can be parameterized by its eccentricity, $e_{c}$. In
Fig.\ \ref{fig:ellipsewaves} (b), the caustic consists of two confocal  
hyperbola segments. 
Caustics separate the classically allowed from 
the forbidden regions, in the sense of the WKB approximation. In this 
section, we discuss some examples of how quasiclassical and exact 
solutions can be obtained in non-chaotic but nontrivial cavities, if 
coupling to the exterior region is neglected. In this closed limit, 
all fields can be written as real-valued functions obeying standard 
boundary conditions (Dirichlet or Neumann). 

\begin{figure}[t]
            \centering
        \includegraphics[width=9 cm]{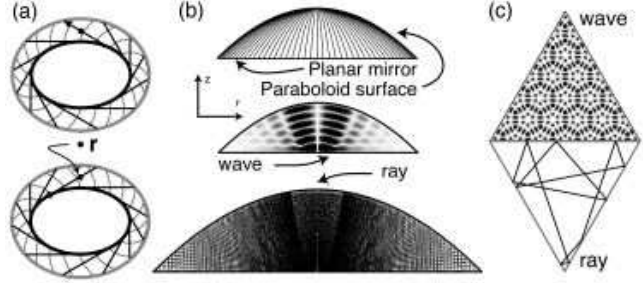}
        \caption{(a) In a WGM of the ellipse, the point $\bf r$ is 
        visited by two wavefronts: one carrying rays from the caustic 
        to the boundary (top), and the other {\em vice versa} 
        (bottom); note that both have the same sense of rotation 
        (indicated by the arrow). A complete ray trajectory 
        alternates between these two wavefronts as it encounters 
        boundary and caustic repeatedly. 
        (b) A paraboloid (surface of revolution generated by a 
        parabola), closed by a mirror intersecting its focal point. The 
        focal length equals $3.25\,\lambda$. The azimuthal mode number
        is $m=1$; high intensity is shown in black. The 
        corresponding rays (bottom) come in 
        families and generally do not close on themselves (bottom). 
        (c) A 2D equilateral triangle cavity. The side 
        length is $13.5\,\lambda$. The mode can be constructed rigorously 
        from a family of periodic ray orbits, one example of which is shown. 
 }
        \label{fig:integrableshapes}
\end{figure}
Figure \ref{fig:integrableshapes} (a) illustrates the WG caustics of 
an ellipse and 
their relation to the quasiclassical quantization method of Einstein, 
Brillouin and Keller (EBK, the multidimensional generalization of the 
WKB approximation) \cite{keller}: given a particular caustic of 
eccentricity $e_{c}$, 
we examine the ray segments that are tangent to it. For a fixed sense 
of rotation, there are exactly two distinct rays going through any 
given point ${\bf r}$ between the caustic and the boundary -- one 
traveling from the caustic toward the boundary and the other {\em 
vice versa}. These two unique ray directions as a function of ${\bf 
r}$ define two vector fields, which are called ``ray congruences'' 
\cite{keller} and are shown in the figure as a flywheel of rays. 
They are furthermore normal to the 
phase fronts shown as curved grey lines. These are the 
eikonals $\Phi_{1,2}({\bf r})$ appearing in the quasiclassical ansatz 
for the wave, 
\be\label{eq:ebkansatz}
\psi({\bf r})=A_{1}({\bf r})\,
\exp[-i\,k\,\Phi_{1}({\bf r})]+
A_{2}({\bf r})\,\exp[-i\,k\,\Phi_{2}({\bf r})]
\ee
We have suppressed the monochromatic time dependence $\exp(i\omega 
t)$. At least two terms are necessary when there are two degrees of 
freedom; additional, symmetry-related terms may be needed to make 
the wave field real-valued. 
In the standard EBK quantization, the amplitude functions $
A_{1,2}$ are assumed to be slowly varying, and one can achieve the 
single-valuedness of the wave function only if the phase advance in 
the exponentials is an integer multiple of $2\,\pi$ for any closed 
loop in the planar cavity. This occurs only for certain discrete 
combinations of the unknown parameters $k$ (the wavenumber), and 
$e_{c}$ (the eccentricity of the caustic). Hence, the semiclassical 
method quantizes not only the wave parameter 
$k$ of the modes, but also the classical parameter $e_{c}$ defining 
the corresponding ray trajectories. 

A simple example is the {\em circular resonator}, a limiting case of 
Fig.\ \ref{fig:integrableshapes} (a). The internal caustic in that 
case is a concentric circle with radius $R_{i}$. By geometry, 
the rays corresponding 
to that caustic have a fixed angle of incidence given by 
$\sin\chi=R_{i}/R$, if $R$ is the cavity radius. Single-valuedness of 
the wave field then requires that the phase advance along a loop 
encircling the caustic ($=$ circumference $\times$ wavenumber) equals 
$2\,\pi\,m$, where $m$ is an integer:
\be\label{eq:angularmomentum}
2\pi\,R_{i}\,nk=2\pi\,m\quad\Rightarrow\quad
\sin\chi=\frac{m}{nkR},
\ee
where $n$ is the index of refraction. The meaning of $m$ follows if 
we 
interpret $p\equiv \hbar nk$ as the linear photon momentum, and 
recall the definition of classical angular momentum, 
${\bf L}={\bf r}\times{\bf p}$. Then if ${\bf r}$ lies on the surface, 
the $z$ component of ${\bf L}$ is $L_{z}=R\,p\,\sin\chi$, which 
identifies $L_{z}=\hbar m$ by comparing with 
Eq.\ (\ref{eq:angularmomentum}). One can thus call $m$ an ``angular
momentum quantum number''. 
In addition to this orbital angular momentum quantization, there is a radial 
single-valuedness condition which forces $k$ to become discrete. This
yields a complete set of two quantum numbers for two degrees of freedom,
and it 
implies that $\chi$ in Eq.\ (\ref{eq:angularmomentum}) becomes
discretized as well -- another way of understanding the quantization of the
caustic $e_c$. For a basic discussion of the EBK 
method in rotationally invariant, separable cavity geometries, 
cf.\ Ref.\ \cite{schweiger}.

One result of the EBK quantization is that the  
WGMs do {\em not} in general correspond to closed ray orbits 
except in the limit when the internal caustic approaches the cavity 
surface. The quantized caustics instead belong to a family of rays 
which encircle the perimeter {\em 
quasi-periodically}, coming arbitrarily close to any given point on 
the boundary after a sufficiently long path length. This 
generalizes to most other resonator problems: 
{\em What counts for the formation of modes is not that the 
associated rays close on themselves, but only that the 
wave fronts, to which these rays are normal, interfere 
constructively}. 

Even in three-dimensional integrable cavities, the EBK method can 
yield highly accurate 
results down to the lowest-frequency modes. As an example, we mention 
recent work on a microlaser cavity with strong internal focusing 
properties, which consists of a dome in the shape of a paraboloid, on 
top of a layered semiconductor \cite{dome}, cf. \
Fig.\ \ref{fig:integrableshapes} (b). As in the ellipse, the exact 
solution for the mode shown for this plano-parabolic mirror geometry 
bears some resemblance to the more familiar paraxial optics, in this 
case a Gauss-Laguerre beam. However, just like the circular 
resonator, the parabolic dome has {\em no} stable ray orbits, owing 
to the confocal condition. The modes can be found exactly because the 
geometry allows separation of variables in parabolic cylinder 
coordinates. 

The short-wavelength approximation can be made highly accurate (as in 
the parabolic dome) or even {\em exact} (as in the Fabry-Perot cavity of 
section \ref{sec:qbstates}). As a nontrivial 
generalization of exact quantization based on rays, 
Fig.\ \ref{fig:integrableshapes} (c) shows 
the {\em equilateral triangle}. 
All its modes can be obtained by superimposing a finite number of 
suitably chosen plane waves \cite{rayleigh,richensberry,laerinoeckelbook}. 
This is achieved by ``unfolding'' the cavity into an infinite lattice 
created by its mirror images.
These examples show that ray optics, as the skeleton which 
carries the wave fields, remains a useful tool far beyond the 
paraxial limit. 

However, the reader may ask: {\em what is the use for 
semiclassical methods in exactly solvable problems such as the above 
systems}? After all, semiclassics is simple in separable systems! 
Beyond quantitative estimates, the value of quasiclassics is that the 
connection between modes and rays can be carried 
over to deformations of the cavity shape where the separability is 
destroyed. When this happens, eigenstates cannot be labeled uniquely 
by global quantum numbers anymore. However, as Weyl's formula teaches 
us, the 
absence of good quantum numbers does not imply the absence of good 
modes. In order to classify the latter, we shall attempt to 
label them according to the ray trajectories to which they 
quasiclassically correspond. 

\section{Chaotic cavities}\label{sec:chaos}
\begin{figure}[t]
            \centering
        \includegraphics[width=12 cm]{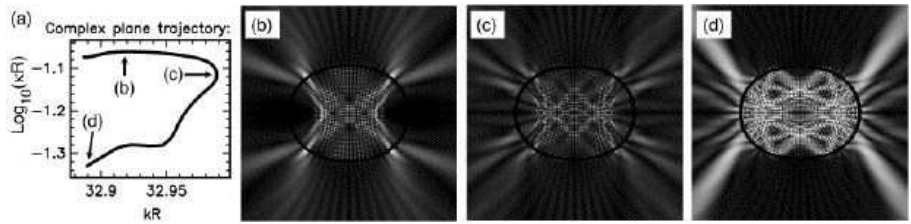}
        \caption{The refractive index in (a-d) is $n=2.65$. 
        (a) Position ${\tilde k}=k+i\kappa$ of a quasibound 
        state, parameterized 
        by the quadrupolar shape deformation 
        $r(\phi)=R\,(1+\epsilon\,\cos 2\phi)/\sqrt{1+\epsilon^2/2}$. 
        Here, $\epsilon$ is the fractional deformation from the circle.
        Arrows show the location of 
        the quasibound states shown in (b-d), corresponding to 
        $\epsilon=0.114$ (b), $\epsilon=0.126$ (c) and 
        $\epsilon=0.135$ (d). The resonance widths are 
        $\kappa\,R=0.087,\,0.069,\,0.047$ in (b), (c), (d). 
        Thus, increasing deformation leads to {\em 
        linewidth narrowing and more focused emission}. 
 }
        \label{fig:chaostransition}
\end{figure}
Taken together, the discussions of stable resonators in the previous 
chapter \cite{siegmanchapter} and of integrable systems in the
foregoing section  
provide the essentials of ``conventional'' resonator physics. However, 
in the infinite space of possible cavity shapes, most geometries are 
nonintegrable and hence display chaos in their ray dynamics, as 
mentioned in the introduction to this chapter. 
To illustrate the transition to chaos, 
Fig.\ \ref{fig:chaostransition} shows the 
continuous evolution of an individual quasibound state as 
the shape of an oval resonator is deformed. 
The cavity shape is a two-dimensional quadrupole of mean radius $R$. 
Now the distinction between the mathematical ellipse and other oval 
shapes becomes important. Although the difference between the shapes 
in Figs.\ \ref{fig:chaostransition} and \ref{fig:ellipsewaves} is 
barely discernible at small $\epsilon$, the quadrupole is not an 
integrable cavity and displays a far more intricate internal mode 
structure. In particular, caustics become frayed, and nodal lines 
form ever more complicated patterns as $\epsilon$ increases. 

Given this complicated scenario, it is not immediately clear that 
{\em ray 
considerations} can help at all in understanding the properties of 
states such as those in Fig.\ \ref{fig:chaostransition}. 
However, it turns out that quite the opposite is true: it is the added 
complexity of the internal ray dynamics that can be identified as 
the {\em cause} of the more complex wave fields. We make this somewhat 
provocative statement because the previous examples have proven the 
success of using ray considerations as a scaffolding for 
constructing the cavity modes. 
Unfortunately, there is so far no complete theoretical framework for 
the quasiclassical quantization of partially chaotic systems; 
the EBK method, which in quantum mechanics gives rise to the 
corrected Bohr-Sommerfeld quantization rules \cite{keller}, fails 
when the system does not 
exhibit well-defined caustics, as was already noted by Einstein 
\cite{einstein1917}. 

However, it is worth following the quasiclassical 
route because there exists a vast amount of knowledge about the 
classical part of the problem: chaotic ray dynamics can 
help gain insights into the wave solutions that cannot be gleaned 
from numerical computations alone. Ray optics can be formally mapped 
onto the classical mechanics of a point particle; this allows us to
leverage a rich body of work on chaotic classical mechanics --
a mature, though by no means complete 
field \cite{reichl,gutzwillerbook,arnold,zaslavsky}. Hence, 
quasiclassics in the presence of chaos is a 
challenging undertaking, but also a {\em useful} one 
because the numerical costs for obtaining exact wave solutions in 
nonintegrable systems are so high.

As observed in section \ref{sec:experiments}, pulses that excite
many cavity modes can behave ballistically, i.e. seem to follow
well-defined trajectories. The idea of the quasiclassical 
approach is complementary to this: from the behavior of whole {\em
families} of ray trajectories, we want to extract the properties of
individual cavity modes. 

\section{Phase space representation with Poincar{\'e} sections}
\label{sec:poincaresections}
As the ellipse already taught us, different types of ray motion 
can {\em coexist} in a single resonator -- e.g., WG and 
bouncing-ball trajectories. In such cases, it is desirable to know 
what combination of initial conditions for a ray will result in 
which type of motion. Initial conditions can be specified by 
giving the position on the boundary at which a ray is launched, 
and the angle $\chi$ it forms with the surface normal. The proper 
choice of initial conditions was identified at the end of section
\ref{sec:qbstates} as a prerequisite in the quasi-classical modeling 
of resonance decay. The present section introduces the tools necessary for 
solving this problem.

\begin{figure}[t]
            \centering
        \includegraphics[width=11 cm]{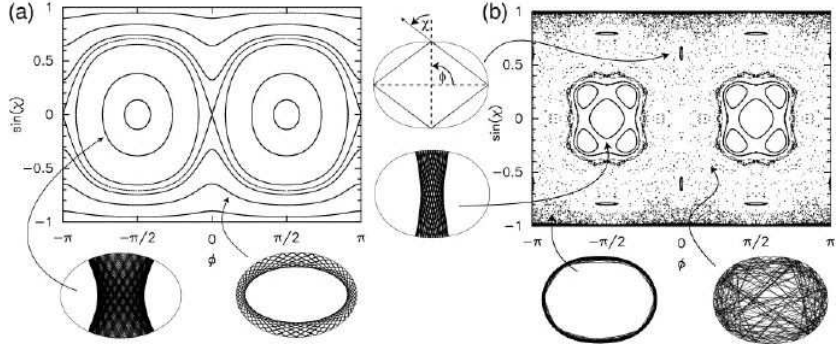}
        \caption{
        Poincar{\'e} surfaces of section: at each reflection, record the 
        position of impact (parameterized by the polar angle $\phi$) and angle 
        of incidence $\chi$, in a plot of $\sin\chi$ versus $\phi$; 
        repeat for several trajectories. 
        (a) ellipse of eccentricity $e=0.8$ (same shape as in 
        Fig.\ \ref{fig:ellipsespec}); (b) quadrupole 
        with $\epsilon=0.135$, as in 
        Fig.\ \ref{fig:chaostransition} (d). In (a),
        oscillatory paths between the flat sides appear as 
        closed loops, whereas WG circulation creates curves that 
        span all polar angles $\phi$. While all rays in 
        (a) form caustics, the chaotic paths shown in (b) do not. 
        Chaotic rays can 
        undergo WG circulation (bottom left) for long times before covering 
        the whole cavity in a quasi-random way (bottom right). 
        }
        \label{fig:sosintro}
    \end{figure}
Information on the possible types of ray motion, and 
how they are grouped in families, is contained in the 
{\em Poincar{\'e} surface of section} (SOS), 
cf.\ Fig.\ \ref{fig:sosintro}. It is a  representation of the classical 
phase space in the $\phi$ - $\sin\chi$ plane, spanning all possible 
angles of incidence and reflection positions along the boundary. 
The definition of $\chi$ in the center of Fig.\ \ref{fig:sosintro} 
includes a {\em sign}, measured positive in the direction shown by the 
arrow from the outward normal. This reflects the observation 
made below Eq.\ (\ref{eq:angularmomentum}) that $\sin\chi$ is 
proportional to the $z$ component of the instantaneous angular momentum 
at the reflection, and the sign thus distinguishes the sense of 
rotation. 

By combining the two variables $\phi$ and $\sin\chi$, dynamical 
structure can be revealed which would remain hidden in a collection of 
real-space ray traces. Non-chaotic trajectories are confined by 
local or global conservation laws to one-dimensional lines, called 
{\em invariant curves}. Almost all rays in the ellipse follow 
such curves, as seen in Fig.\ \ref{fig:sosintro} (a), which shows a 
clearcut division into oscillatory  and rotational (WG) 
motion, cf.\ also Fig.\ \ref{fig:ellipsewaves}. This is analogous to 
the phase space of a physical pendulum \cite{goldstein}. The separatrix 
between the two types of motion corresponds to a diametral ray orbit 
connecting the points $\phi=0,\,\pi$ on the boundary. 

Rays launched on an invariant curve must remain 
on it for all subsequent reflections. Such unbroken curves persist 
for $|\sin\chi|\to 1$ even in Fig.\ \ref{fig:sosintro} (b); this is 
just the WG limit.
A lesser degree of robustness is observed for the oscillatory 
trajectories: they are surrounded by elliptical ``stable islands'', 
but a chaotic sea forms in-between, owing to the fact that 
in Fig.\ \ref{fig:sosintro} (b) the diametral separatrix orbit 
mentioned above is {\em unstable}, developing the 
sensitivity to small deviations in initial conditions which is typical 
for chaotic behavior. Chaos develops 
preferentially around such separatrix orbits. 
The analogy to a pendulum makes this
plausible: there, the separatrix is the unstable 
equilibrium point at which the pendulum balances upside down. 
Note that the different types of motion described
here are {\em mutually exclusive}, i.e. chaotic orbits never 
cross over into the islands of stability. This has the important
effect that chaotic motion is indirectly affected by the presence of
stable structure in the SOS. 

The central reason why the SOS is introduced in this 
chapter is that it allows us to form a bridge between the discrete 
electromagnetic modes of the cavity and their measurable emission 
characteristics. Emission means coupling to the environment and hence 
appears as a dissipation mechanism in the internal cavity dynamics, as 
illustrated in Eq.\ (\ref{eq:raydecay}). Recalling section 
\ref{sec:dielectrics}, the emission from a 
dielectric cavity is governed foremost by Fresnel's formulas, which 
give a wavelength-independent relation for the reflectivities 
$r_{\nu}$ along any ray path, determined only by the local angle of 
incidence $\chi_{\nu}$ at reflection $\nu$; in 
the example of the Fabry-Perot cavity, nothing else is needed to 
determine the decay rate 
$\gamma$ of a cavity mode, cf.\ Eq.\ (\ref{eq:gammareflect}). 
Now we note that the SOS shows the angle of incidence on the vertical 
axis, and hence the Fresnel coefficient of any given reflection can 
be read off with ease. In particular, the critical angle for TIR, 
Eq.\ (\ref{eq:criticalangle}), is represented in 
the SOS as a horizontal line. When this line is crossed from above (in 
absolute value), refractive escape becomes possible. The SOS then 
tells us which combinations of initial conditions give rise to ray 
trajectories that eventually reach this classical escape window in 
phase space. 

\section{Uncertainty principle}
To complete the bridge between a cavity mode and its 
corresponding set of initial conditions in the ray dynamics, we have 
to find a way of projecting quasibound fields onto the SOS. 
To this end, let us briefly discuss some aspects of how mode fields 
can be measured. 
Individual quasibound states can be studied in great detail in 
{\em microlaser} experiments \cite{mekis,gmachl,laeri,sschang},
because one can make spatially and spectrally 
resolved images of the emitter under various observation angles. As 
is evident from Fig.\ \ref{fig:chaostransition}, the far-field intensity 
depends on the polar angle $\theta$ of the detector 
relative to some fixed cross-sectional axis of the object (say, the 
horizontal axis). Instead of simply measuring this 
far-field intensity, however, one can also ask from which points on 
the cavity surface the collected light originated -- i.e., an {\em 
image} of the emitter can be recreated with the help of a lens. The 
dielectric will then exhibit bright spots at surface locations $\phi$ 
whose distribution in the image may change as a function of observation 
angle $\theta$. The two variables $\phi$ and $\theta$ are 
conjugate to each other, because $\theta$ measures a propagation {\em 
direction} whereas $\phi$ is a position coordinate of the cavity. 

The conjugacy between $\phi$ and $\theta$ means 
that they are {\em incompatible}, in the sense that they 
obey an {\em uncertainty relation}: 
the field distribution as a function of $\phi$ in the image can be deduced 
from the far-field distribution in $\theta$ by a Fourier 
transformation, and that is the function performed by 
the imaging lens. But quantities related by Fourier transformation 
cannot simultaneously have arbitrarily sharp distributions. 
Physically, in an image-field measurement a large lens is needed to get 
good spatial ($\phi$) resolution, but this leaves a larger uncertainty 
about the direction $\theta$ in which the collected light was traveling 
\cite{sschang}. 

If we could plot the measured intensity as a function of both 
$\phi$ and $\theta$, we could generate a two-dimensional distribution 
similar to the SOS of the preceding ray analysis. In 
fact, knowing the surface shape, it is only a matter of trigonometry 
and the law of refraction to transform the pair $(\phi,\,\theta)$ 
measured on the outside to 
$(\phi,\,\chi)$ inside the cavity and hence make the analogy complete. 
But if one has 
already calculated the wave field of a quasibound state, what is the 
advantage of representing it in this {\em phase space} rather than in 
real space? The answer is that the real-space wave patterns often 
obscure information about the {\em correlations} between the two 
conjugate variables $(\phi,\,\sin\chi)$. As mentioned above, the SOS 
which plots these variables would allow us to understand whether a given 
mode is allowed to emit refractively; and if so, we can furthermore 
determine at what positions on the surface and in which directions the 
escape will preferentially occur. This is determined by the joint 
distribution of $(\phi_{\nu},\,\sin\chi_{\nu})$ over all 
reflections $\nu$ encountered by the rays corresponding to the given 
mode \cite{nature}. 

\section{Husimi projection}
To begin a phase space analysis, we now return to the question of 
how to extract correlations between conjugate variables from a wave 
field. This can be illustrated by analogy with a musical score: 
any piece of music can be recorded by graphing the 
sound amplitude as a function of time. On the other hand, musical 
notation 
instead plots a sequence of sounds by simultaneously specifying 
their pitch and duration. These are conjugate variables because 
monochromatic sounds require infinite duration \cite{bartelt}, but 
their joint distribution is what makes the melody. 
The reason why musical scores can be written unambiguously is that they 
apply to a short-wavelength regime in which the frequency smearing by 
finite ``pulse'' duration goes to zero (in this sense, all music is 
``classical''\ldots). In the same way that musical notes are adapted 
to our perception of music, phase-space representations of optical 
wave fields project complex spatial patterns onto classical variables 
relevant in measurements. 

This reasoning is familiar from quantum optics as well: there, 
amplitude and phase of the electromagnetic field are conjugate 
variables, and 
their joint distribution is probed in correlation 
experiments \cite{walls,raymer}. A possible way of obtaining a 
phase-space 
representation is the {\em Husimi function}, obtained by forming an 
overlap integral between the relevant quantum state and a {\em 
coherent state} corresponding to a minimum-uncertainty wavepacket in 
the space of photon number versus phase. Here, we want to extract the 
same type of information about the joint 
distribution of the conjugate variables $\phi,\sin\chi$ relevant to 
an optical imaging measurement on a cavity mode, as described above. 
By projecting the electromagnetic field onto the SOS, measurable 
correlations are revealed which can be compared to the 
classical phase space structure. This is another realization of a 
Husimi function; the examples mentioned above differ essentially only 
in the actual definition of the coherent state basis onto which 
the wave field is projected \cite{leboef}.

Motivated by the above remarks on measurement of emission locations 
versus detector direction, we now note that the coherent 
states with which the cavity modes should be overlapped are in 
fact the {\em Gaussian beams} \cite{siegman}. The fundamental 
Gaussian beam $\psi_{G}$ has the 
property of being a minimum-uncertainty wavepacket in the coordinate $x$ 
transverse to its propagation direction $z$, evolving according to 
the paraxial (Fresnel) approximation in complete analogy with a 
Schr{\"o}dinger wavepacket,
\be\label{eq:minumumuncert}
\psi_{G}(x,z)=\frac{1}{(2 \pi)^{1/4}}\,\frac{1}{\sqrt{\sigma+\frac{i\,z}{2 
k\sigma}}}\,\exp\left[-\frac{x^2}{4\sigma^2+2iz/k}\right]\,\exp(ikz).
\ee
The ``waist'' of the beam at $z=0$ is $\sigma$, and the 
{\em angular beam spread} is 
\be 
\Delta\theta=\arctan\left(\frac{1}{k\,\sigma}\right)
\approx\frac{1}{k\,\sigma}.
\ee
In the imaging setup, $\sigma$ determines the spatial resolution. 
Only beams with angular spread smaller than a maximum $\Delta\theta$ 
are admitted by the aperture, and hence features smaller than the 
corresponding $\sigma$ are unresolved. Clearly, the uncertainty 
product $\sigma\,\Delta\theta$ vanishes for high wavenumbers. 

We are concerned with the 
{\em internal} cavity, and thus do not want the definition to 
contain the particular leakage mechanism (e.g., the law of 
refraction). Therefore, imagine that our detector could 
be placed {\em inside} the cavity, close to its boundary. Given the 
internal field $\psi_{{\rm int}}$ of the mode, we then make a 
hypothetical measurement by forming the overlap with a 
minimum-uncertainty wavepacket $\psi_{G}$ of width $\sigma$, centered 
around a certain value of $\phi$ and $\sin\chi$. The form of $\psi_{G}$ 
is analogous to Eq.\ (\ref{eq:minumumuncert}), but transformed 
to polar coordinates because our position variable is an 
angle, $\phi$. The radial coordinate can be eliminated because we 
constrained our detector to lie on the cavity surface. 

Although individual Gaussian wavepackets are not good solutions of 
the wave problem inside the cavity, they are {\em always} an 
allowed (though overcomplete) basis in which the field can be 
expanded. This is all we require in order to obtain the desired, 
smoothed phase space representation of $\psi_{{\rm int}}$:  
The value of the overlap integral, 
\be\label{eq:husimidef}
H(\phi,\sin\chi)\equiv \left|\langle 
\psi_{int}|\psi_{G}\rangle\right|^2,
\ee
as a function of $\phi$ and 
$\sin\chi$ can (with some caveats) be interpreted as a phase space 
density, and is the desired Husimi projection; the parameter $\sigma$ 
in Eq.\ (\ref{eq:minumumuncert}) should be chosen so as to optimize 
the desired resolution in $\phi$ and $\sin\chi$, i.e., the 
``squeezing'' of the minimum-uncertainty wavepackets which probe 
$\psi_{int}$. Details on the definition, properties and applications 
of Husimi functions on the ray phase space are found in 
Refs. \ \cite{smilansky,crespi,tualle,frischat,
hackenbroich,gmachl}. The price we pay for obtaining a 
quasi-probability distribution is that Eq.\ (\ref{eq:husimidef}) 
discards all phase information. 

\begin{figure}[!t]
            \centering
        \includegraphics[width=13cm]{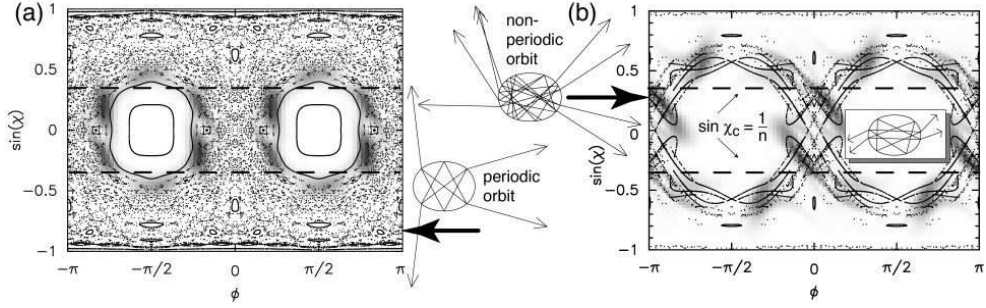}
        \caption{The Husimi projections of states (b) and (d) in 
        Fig.\ \ref{fig:chaostransition} are shown in (a) and (b), resp., as 
        grayscale images (dark for high intensity) 
        superimposed onto the SOS. In (a), the field is 
        localized in the chaotic boundary 
        layer of the bouncing-ball island, with strong maxima on an 
        unstable periodic orbit. Its star shape is shown in the bottom 
        center inset, displaying ray escape according to Snell's 
        law is as thin arrows. In (b), the cavity 
        is the same as in Fig.\ \ref{fig:sosintro} (b), but the SOS shows 
        a different set of ray trajectories, forming 
        a homoclinic tangle (interweaving black lines). The rightmost inset 
        shows the unstable periodic orbit from which the tangle emanates. 
        Its invariant manifolds guide the transport of 
        phase-space density to the refractive escape window, 
        $\vert\sin\chi\vert<1/n\approx 0.377$ (dashed 
        horizontal lines). 
 }
        \label{fig:husimi}
\end{figure}

Figure \ref{fig:husimi} shows how the prescription Eq.\ (\ref{eq:husimidef})
maps the mode of Fig.\ \ref{fig:chaostransition} (c) and (d) onto the 
classical ray dynamics. Although the chaotic trajectories 
corresponding to (d) show some 
accumulation near certain paths in real space, only the Husimi plot 
reveals what {\em combinations} of $\phi$ and $\sin\chi$ occur in the 
internal field. This in turn determines at what positions refractive 
ray escape is expected, and in which directions the light will radiate. 

\section{Constructive interference with chaotic rays}
\label{sec:chaoticinterference}
In this section, we return to the ``mystery of the missing modes'' with 
which section \ref{sec:resoqbstates} confronted us: Weyl's density of 
states is asymptotically the same for all cavities of the same volume, 
even if we cannot classify their modes according some numbering that 
follows from integrability, or at least paraxiality. 
Figure \ref{fig:chaostransition} shows that a given mode 
does not cease to exist as integrability is destroyed with increasing 
$\epsilon$, but evolves smoothly into a new and complex field 
pattern with a well-defined spectral peak that can even become 
narrower with increasing chaos in the ray dynamics. As noted in 
section \ref{sec:chaos}, the ``mystery'' 
is how at large shape deformation the increasingly numerous chaotic 
rays, which crisscross the 
cavity in a pseudo-random way and even diverge from each other, can 
give rise to constructive interference, and hence to a resonant mode. 

In section \ref{sec:integrableshapes}, the quasiclassical quantization 
approach led from the ray picture to the wave fields. Now, 
we follow the opposite direction and use the Husimi projection of a 
given cavity mode to perform a ``dequantization'' which leads to the 
ray picture. In this way, the novel physics of chaotic modes can be 
described without much technical detail.

By superimposing the Husimi projection of a mode onto the SOS
for the same deformation, we can identify classical 
structures on which the mode is built. We have so far encountered two 
types of structure, cf.\ the caption of Fig.\ \ref{fig:sosintro}:
{\em (I)} invariant curves such as those formed by WG rays, permitting EBK 
quantization as discussed in section 
\ref{sec:integrableshapes}; {\em (II)} stable islands around periodic orbits, 
for which the paraxial approximation of the previous chapter 
\cite{siegmanchapter} can be used. In both cases, a finite (and usually small) 
number of wavefronts suffices to achieve the constructive interference 
without which no cavity mode can form. 
The mode shown in Fig.\ \ref{fig:husimi}, on the other hand,  
overlaps with neither of these non-chaotic phase-space components; it is 
a {\em chaotic} mode. One possible way to understand that chaotic 
modes are possible at all, is to realize that even the chaotic sea is 
{\em not} structureless. 

The 
classical objects of crucial importance for an 
understanding of chaotic modes are the {\em unstable} periodic orbits 
of the cavity. They give rise to a third type of invariant sets in the 
ray dynamics: {\em (III) stable and unstable manifolds}. These are 
one-dimensional curves which, like the structures {\em (I)} and 
{\em (II)}, have 
the property that rays launched anywhere on the manifold will always 
remain there. Along these manifolds, trajectories rapidly 
approach or depart from a given periodic orbit, such as the one shown 
in Fig.\ \ref{fig:husimi} (b, right inset).  A 
non-periodic ray trajectory which moves along one of the corresponding 
manifolds is displayed in the top center inset. 

The stable and unstable manifolds form an intricate web, 
the ``homoclinic 
tangle'', which was already recognized by Poincar{\'e} as the cause of
severe difficulties in calculating the dynamics 
\cite{gutzwillerbook}. Part of this tangle is shown in 
Fig.\ \ref{fig:husimi} (b), forming interweaving lobes which play an 
important role in controlling the transport of phase-space density.
A discussion of this ``turnstile'' action and the relation 
to the stability of the orbits from which the tangle originates is 
found, e.g., in Refs.\ \cite{reichl,bohigas}). 
As the Husimi projection in 
Fig.\ \ref{fig:husimi} (b) shows, the wave intensity is guided 
along the invariant manifolds of the unstable periodic orbit \cite{uzer}, 
imparting a high degree of anisotropy onto the internal intensity and 
on the emission directions. 

Particularly strong Husimi intensity is
found at the reflection points of the periodic orbit shown in the inset 
to Fig.\ \ref{fig:husimi} (b); this corresponds to the high-intensity 
ridges in the real-space intensity 
of Fig.\ \ref{fig:chaostransition} (d). 
Such {\em wavefunction scarring} 
\cite{heller} is a surprising feature, because 
a wave field concentrated near the origin of the homoclinic tangle, 
i.e., at the unstable periodic orbit itself, spatially appears similar 
to a sequence of Gaussian beams, despite the fact that 
paraxial optics requires the underlying modes to be {\em stable}. Even 
if we do not start from this paraxial point of view, but follow the 
historical developments in quantum chaos, scars seem to run counter 
to the early conventional wisdom that asserted chaotic modes should 
posses a random spatial distribution \cite{bohigas}. 
In a recent study on lasing in a large, planar laser cavity with 
no stable ray orbits \cite{fukushima}, laser operation with focused 
emission 
was observed and explained by wave function scarring. This phenomenon 
is of interest not only from the applied point of view, but also 
because a complete understanding of the quasiclassical theory for 
modes corresponding to the invariant manifolds of type {\em (III)} 
is as yet missing \cite{tomsovichellerconstr,kaplan}.  

\section{Chaotic whispering-gallery modes}\label{sec:chaowgmodes}
\begin{figure}[!t]
            \centering
        \includegraphics[width=13 cm]{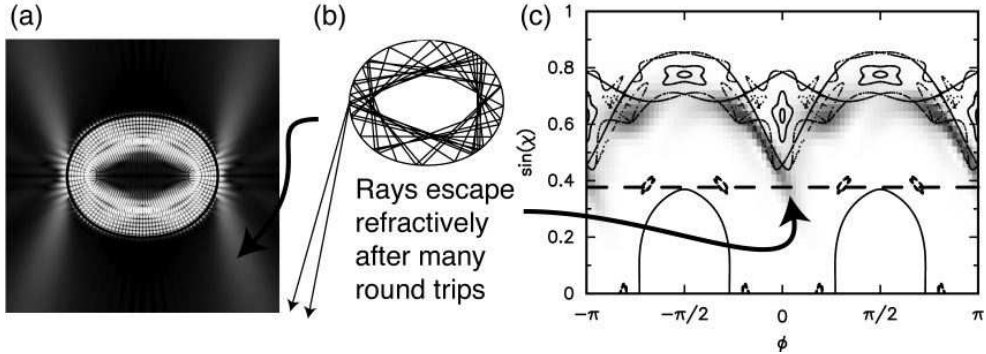}
        \caption{A WGM in the quadrupole cavity 
        at $\epsilon=0.104$, at dimensionless wavenumber $kR=32.695$ and 
        width $\kappa R=0.0025$. The focused near-tangential emission 
        from the points of highest curvature is explained in the ray 
    picture (b). The ray-wave connection is illustrated in the Husimi 
    plot (c), superimposed onto the SOS (only $\sin\chi>0$ is plotted 
    to magnify details). The stable and unstable 
    manifolds of the rectangular periodic orbit funnel the rays 
    downwards to the escape condition along two narrow pathways. The 
    resulting emission therefore is not attributable to a single 
    periodic orbit, but to a family of (generally aperiodic) 
    trajectories following the invariant manifolds shown here. 
 } \label{fig:wghusimi}
\end{figure}
The lifetime of the quasibound states in 
Fig.\ \ref{fig:chaostransition} (d) yields a Q-factor of $Q\equiv 
k/\kappa=700$, which is of the same order as that of a high-power 
microlaser \cite{gmachl} that was demonstrated recently as an 
application of cavity design using the concepts introduced in section
\ref{sec:poincaresections}. The bowtie-shaped mode encountered there 
was of paraxial type, centered around a stable periodic ray orbit. If 
low lasing thresholds and hence higher Q-factors are desired, we 
recall from Fig.\ \ref{fig:ellipsewaves} that WGMs are  
preferred. As shown in Fig.\ \ref{fig:wghusimi}, WGMs in partially 
{\em chaotic} cavities display high Q {\em and} strongly anisotropic 
emission patterns. Comparing the non-chaotic case of 
Fig.\ \ref{fig:ellipsewaves} (a) and the chaotic cavity of 
Fig.\ \ref{fig:wghusimi}, we indeed find stronger focusing in the
latter case. 

Chaotic WGMs are
among the few types of chaotic modes for which an approximate
quasiclassical description along the lines of section 
\ref{sec:integrableshapes} is possible. 
They correspond to ray trajectories which 
are part of the chaotic sea but circulate along the cavity perimeter 
many times before exploring other phase-space regions, 
cf.\ Fig.\ \ref{fig:sosintro} (b), bottom. There is thus a 
{\em separation of time scales} between the fast WG circulation 
and a slow deviation from WG behavior which inevitably ends in chaotic 
motion. This makes it possible to use 
an adiabatic approximation \cite{sweden} in which 
the ``diffusion'' away from WG motion is neglected for times
long enough to contain many round-trips of the ray. 

As a
consequence, one can formulate approximate quantization conditions for
chaotic WG modes by ignoring the chaotic dynamics \cite{nature}, 
leading to equations
of the EBK type, Eq.\ (\ref{eq:ebkansatz}). The approximate nature of
this quantization can be recognized in Fig.\ \ref{fig:wghusimi} (a), 
which differs from the ellipse-shaped cavity of 
Fig.\ \ref{fig:ellipsewaves} (a) in the fact that a small but 
nonvanishing field persists in the cavity 
center. This indicates that there no longer is a well-defined caustic, 
as we required in section \ref{sec:integrableshapes} Nevertheless, 
the Husimi distribution (grayscale) in Fig.\ \ref{fig:wghusimi} (c)
condenses approximately onto a one-dimensional curve (the shape of
which can be given analytically within the adiabatic approximation
\cite{nature}), following the homoclinic tangle [the SOS in  
Fig.\ \ref{fig:wghusimi} (c) is vertically expanded compared to 
Fig.\ \ref{fig:husimi} (b)]. The location of this 
{\em adiabatic invariant curve} is determined by the EBK
quantization; its minima are seen to approach the TIR condition.

As indicated in section \ref{sec:chaoticinterference},
chaos does become important when the emission properties 
of a chaotic WG mode are concerned. 
It is the deviation of the rays from the above
adiabatic assumption which allows an initially confined ray to violate
the total-internal reflection condition after some time, and hence
escape refractively, cf.\ Fig.\ \ref{fig:wghusimi} (b). In 
Fig.\ \ref{fig:wghusimi} (c), this escape results from wave
intensity leaking across the TIR condition near 
$\phi=0,\,\pi$. This can be reproduced within a ray
simulation by launching an ensemble of rays on the adiabatic invariant
curve, and recording the distribution of classically escaping rays. 

This implements the program outlined at the end of 
section \ref{sec:qbstates}
The stochastic branching of a ray trajectory into 
transmitted and reflected parts, also known as ray-splitting 
\cite{bluemelott}, could be summed up as a geometric series in the
Fabry-Perot example, Eq.\ (\ref{eq:raysum}). In the oval cavity, 
Monte-Carlo simulations for large 
ensembles of incident rays are required \cite{nature}. This ray 
approach has been applied in scattering problems
along the lines of Eq.\ (\ref{eq:miller}) \cite{traenkle}; the value
and applicability of ray considerations in scattering from 
{\em chaotic} optical cavities was demonstrated in 
Ref.\ \cite{jensen}. The difference to the {\em quasibound-state} 
approach is that our incident rays are launched inside the cavity, not
outside of it. 

Because chaotic diffusion makes classical escape from WGMs possible in 
the first place, the ray approach in fact becomes more useful at
large deformations where chaos increases. That is just the regime
where exact scattering calculations encounter difficulties. Starting
from the ray approach to emission directionality and decay rates, one
can then successively introduce corrections at higher orders in
$\lambda$. In particular, in Ref.\ \cite{nature}, the openness of 
the system was taken into account in the semiclassical quantization 
step by introducing additional phase shifts into the EBK
equations. The reason is that penetration into the outside region, 
as mentioned in section \ref{sec:dielectrics}, creates an enlarged, 
effective cavity boundary which lengthens the ray path lengths. This
``Goos-Haenchen effect'' \cite{snyderlove} is well-known from plane 
interfaces as well. The effective cavity is actually of slightly 
different shape than the physical geometry. This has the 
consequence that {\em even the elliptic} cavity of 
Fig.\ \ref{fig:ellipsewaves}, which for impenetrable boundaries is
integrable, becomes non-integrable as a 
dielectric body \cite{yeh,lockellipse}. 

\section{Dynamical eclipsing}
\label{sec:experimentdynec}
In the $\lambda\to 0$ limit, the internal dynamics of the ellipse
displays no chaos, cf.\ Fig.\ \ref{fig:sosintro} (a). This makes it an
important test case with which to compare whenever we want to identify
fingerprints of chaos. The power of such comparisons has been 
illustrated in an experiment on prolate, dye-doped lasing microdroplets
\cite{sschang}. The geometry of the droplets was rotationally symmetric
around the polar axis, which is also the major axis. If one analyzes
only the light propagating in a plane containing the two poles, 
then the cavity is effectively formed by the 2D axial cross section
whose shape is an oval of the type we have been discussing above,
cf.\ Fig.\ \ref{fig:sosintro}. 
The experimental technique of recording with a CCD camera an image of
the micro-object at various observation angles $\theta$ (measured 
with respect to the polar axis of a spheroidal droplet) reveals
prominent bright and dark regions along the cavity rim, from which the
position and exit angle of the laser emission can be extracted
simultaneously. 

As discussed in section \ref{sec:mie}, the lasing modes in
low-index materials such as droplets should be of
WG-type. The experiment illustrates both the
focusing property, as well as the robustness of WGMs. 
When the lasing droplets are imaged from the side
perpendicular to the polar axis ($\theta=90^{\circ}$), 
one thus expects emission directionality of the type shown in 
Figs.\ \ref{fig:ellipsewaves} (a) or \ref{fig:wghusimi} (a), showing 
emission from the points of highest curvature (i.e., the
poles). However, {\em neither} case describes the observed emission
correctly. The droplets appeared dim at the poles instead of being
brightest there. 

This counter-intuitive phenomenon can be explained by the 
phase space structure shown in Fig.\ \ref{fig:sosintro} (b). An
essential role is played by the small islands belonging to the stable 
four-bounce diamond-shaped orbit depicted in the Figure. Its effect is
that WG rays of the type shown below the SOS are prevented from
reaching the high-curvature points ($\Phi=0,\,\pi$) at angle of
incidence $\sin\chi=0.65$, because chaotic and stable motion are
mutually exclusive (cf. section \ref{sec:poincaresections}). Now 
$\chi_{\rm TIR}=\arcsin 0.65$ just happens to be the critical angle for the
droplet-air interface; hence, near-critical escape from the
highest-curvature points is ruled out for chaotically moving WG 
rays in the droplet. In Fig.\ \ref{fig:sosintro} (b), it is impossible
to see from the ray trace in real space that the combination 
$\phi=0$, $\chi=\chi_{\rm TIR}$ will never occur. Likewise, this information
is by no means obvious from the wave equation itself Only the
phase-space analysis using the SOS 
reveals the small islands that cause this effect, which has been
called {\em dynamical eclipsing}. The emission profile is determined
by the phase space structure in the vicinity of the emission line
$\sin\chi_{TIR}$, and therefore we do not detect dynamical eclipsing
in the higher-index example of Fig.\ \ref{fig:wghusimi}. 
A glance at Fig.\ \ref{fig:sosintro} (a)
makes it clear that the effect will never occur in the ellipse, 
because the relevant islands do not exist there. 

\section{Conclusions}
The subtle difference in shapes
between ellipsoid and quadrupole raises the question: {\em How short does
the wavelength have to be compared to the cavity dimensions, in order
to be able to resolve such differences in geometry?} Clearly, the
droplets are so large  compared to the optical wavelength 
($35\,\mu$m diameter for an equivalent-volume sphere) that we are deep
in the short-wavelength limit. What the above example teaches
us is that it is not the structure in real space that has to be 
resolved by the
waves, but the structure in phase space. The uncertainty
principle will limit the size of the islands that can have a noticeable
effect on the cavity mode structure, but the important thing to
realize is that small differences in cavity shape can lead to large
differences in phase space structure. As a result, the phase space
effects introduced in this chapter turn out to be observable
even in cavities that are not much larger than the wavelength. This is
an extension of the claim made for integrable systems in 
section \ref{sec:integrableshapes}, that quasiclassical methods 
often extend all the way to the longest-wavelength modes. 

Future work on nonintegrable dielectric cavities can proceed in various 
directions: our understanding of the {\em classical} ray transport in 
partially chaotic systems has to be explored further, and the 
systematic {\em wave} corrections to these ray ideas must be 
investigated. Two such corrections which are of particular importance 
in chaotic systems are dynamical localization \cite{sweden,nature,dynloc} 
and chaos-assisted tunneling \cite{hackenbroich}. Dynamical
localization is an interference effect that suppresses decay rates
below the value expected from ray simulations as outlined in this
chapter. It becomes more pronounced in smaller cavities, and in fact 
helps maintain high Q-factors in the presence of chaotic
ray dynamics. Localization arises in many areas of
physics as a name for wave corrections to a
classical diffusion picture. In our case, the diffusion happens in the
chaotic phase space of the ray dynamics. 
Chaos-assisted tunneling, on the other hand, 
acts to enhance radiation losses of {\em non-chaotic} WGMs at small 
deformations when chaos pervades only the low-$\sin\chi$ regions of
the SOS. It also leads to a coupling between WG wavefronts
encircling the cavity in counterclockwise and clockwise sense of
rotation, e.g., between the extreme bottom and top of the SOS in 
Fig.\ \ref{fig:chaostransition} (a). In the ray picture, spontaneous
reversal of rotation direction is forbidden for WG trajectories on
invariant curves; the mechanism at work here is tunneling in phase
space. 

The phase space approach to microcavity electrodynamics 
provides a means of classifying the wide variety of 
modes in cavities where ``good quantum numbers'' do not uniquely 
enumerate the spectrum. Semiclassical methods provide the connection
between individual modes and manifolds in phase space, such as the
caustics in section \ref{sec:integrableshapes}, belonging to invariant
curves in the SOS. Open questions remain about how to establish 
this connection in the general case when caustics are broken up 
by ray chaos. The particularly important case of WGMs, however, is
amenable to approximate treatments which permit quantitative
predictions. Many systems in which strong matter-field coupling
enables quantum-electrodynamic studies, are actually large enough to be
in the quasiclassical regime: examples are semiconductor domes 
\cite{dome} (where the effective wavelength is shortened by a high
refractive index), or silica microspheres (cf.\ 
Ref. \cite{haroche,kurizki} and citations therein) where 
atoms close to the dielectric interface interact evanescently with 
WGMs whose angular momentum according to Eq.\ (\ref{eq:angularmomentum})
is $m\sim 10^3$. At all levels of 
microcavity optics -- quantum or classical, linear or nonlinear -- 
the fundamental task is to establish a modal basis, and the methods 
we sketched in this chapter accomplish this in a way that 
provides physical insight. 

\end{document}